\documentclass[prl,aps,twocolumn,footinbib,superscriptaddress,notitlepage]{revtex4-1}
\usepackage{amsmath}
\usepackage{amssymb}
\usepackage{graphicx}
\usepackage{dcolumn}
\usepackage{enumerate}
\usepackage{bm}
\usepackage{xcolor}
\usepackage{comment}
\usepackage{subfigure}
\usepackage{breqn}
\usepackage{hyperref}
\hypersetup{colorlinks=true, citecolor=blue, urlcolor=blue, linkcolor=blue}

\makeatletter
\let\cat@comma@active\@empty
\makeatother

\begin{document}
\title{Photon-mediated correlated hopping in a synthetic ladder}
\author{Anjun Chu}
\email{anjun.chu@colorado.edu}
\affiliation{JILA, NIST and Department of Physics, University of Colorado, Boulder, Colorado 80309, USA}
\affiliation{Center for Theory of Quantum Matter, University of Colorado, Boulder, Colorado 80309, USA}
\author{Asier Pi\~{n}eiro Orioli}
\affiliation{JILA, NIST and Department of Physics, University of Colorado, Boulder, Colorado 80309, USA}
\affiliation{Center for Theory of Quantum Matter, University of Colorado, Boulder, Colorado 80309, USA}
\author{Diego Barberena}
\affiliation{JILA, NIST and Department of Physics, University of Colorado, Boulder, Colorado 80309, USA}
\affiliation{Center for Theory of Quantum Matter, University of Colorado, Boulder, Colorado 80309, USA}
\author{James K. Thompson}
\affiliation{JILA, NIST and Department of Physics, University of Colorado, Boulder, Colorado 80309, USA}
\author{Ana Maria Rey}
\affiliation{JILA, NIST and Department of Physics, University of Colorado, Boulder, Colorado 80309, USA}
\affiliation{Center for Theory of Quantum Matter, University of Colorado, Boulder, Colorado 80309, USA}
\date{\today}

\begin{abstract}
We propose a new direction in quantum simulation that uses multilevel atoms in an optical cavity as a toolbox to engineer new types of bosonic models featuring correlated hopping processes in a synthetic ladder spanned by atomic ground states. 
The underlying mechanisms responsible for correlated hopping are collective cavity-mediated interactions that dress a manifold of excited levels in the far detuned limit. 
By weakly coupling the ground state levels to these dressed states using two laser drives with appropriate detunings, one can engineer correlated hopping processes while suppressing undesired single-particle and collective shifts of the ground state levels.
We discuss the rich many-body dynamics that can be realized in the synthetic ladder including pair production processes, chiral transport and light-cone correlation spreading. 
The latter illustrates that an effective notion of locality can be engineered in a system with fully collective interactions. 
\end{abstract}

\maketitle

{\it Introduction.}---Cavity QED systems are emerging as leading platforms for quantum simulation of tunable long-range interacting spin models and spin-boson models featuring rich steady-state and non-equilibrium many-body behaviors \cite{mivehvar2021cavity,baumann2010dicke,klinder2015observation,landig2016quantum,leonard2017supersolid,Vaidya2018,norcia2018cavity,davis2019photon,muniz2020exploring,schuster2020,zhang2021observation,guo2021optical,periwal2021programmable,konishi2021universal}.
In these systems, impressive progress has been achieved in engineering collective interactions between effective two-level or three-level systems with all-to-all or programmable connectivity \cite{Vaidya2018,periwal2021programmable}. 
Nevertheless, a fascinating avenue that remains yet to be explored is the rich physics emerging with multilevel atoms, which offers new opportunities including dark states immune to cavity decay, subradiant states in free space, and multilevel squeezed states \cite{RitschPRL118, Asenjo_PNAS2019, Orioli_PRL123,JessenDeutsch_PRA2021,orioli2022emergent}.

In this work, we propose the use of multilevel atoms in an optical cavity as a toolbox to engineer different types of bosonic models featuring \emph{correlated} hopping processes, i.e. a process whereby the hopping rate of an atom depends on the presence of other atoms in the array. 
Correlated hopping processes are believed to not only enrich the many-body physics in extended Hubbard models \cite{dutta2015non,meinert2016floquet}, but also play a fundamental role in quantum simulation of dynamical gauge fields \cite{gorg2019realization,schweizer2019floquet} and topological materials \cite{kraus2013majorana,Chanda2021selforganized}.
The key idea is to treat the internal levels of the atoms as a synthetic dimension \cite{celi2014synthetic}, where a notion of spatial locality can naturally emerge, even though photon-mediated interactions are all-to-all in a cavity. 
The use of internal levels as a synthetic spatial dimension has already lead to beautiful demonstrations of topological lattice models and observations of chiral transport in non-interacting models \cite{ozawa2019topological,mancini2015observation,stuhl2015visualizing,kolkowitz2017spin,chalopin2020probing}, and very recently in interacting many-body systems \cite{zhou2022observation}.  
In our case, we propose a way to go beyond the single-particle paradigm by engineering  interaction-induced hopping processes in the synthetic dimension spanned by the atomic ground state manifold. This is accomplished by weakly coupling it to a set of many-body excited states dressed by photon-mediated interactions.

We propose to use two laser drives with appropriate detunings to suppress undesirable single-particle and collective shifts of the internal levels. In this way, we select only the desired hopping process where in a correlated manner one atom moves \emph{two} internal levels up while another atom in the array moves \emph{two} levels down.
The correlated hopping processes we introduce split the ground state manifold into two sets of levels, which we visualize as a synthetic two-leg ladder. 
Here, we study a variety of many-body phenomena that can be realized in this system, including dynamical phase transitions in pair production processes, chiral transport that is tunable via laser detunings and initial state preparation, and correlation spreading and emergent light-cone transport in the synthetic ladder. 
We also discuss a feasible experimental implementation using long-lived alkaline earth atoms. 

\begin{figure}[t]
    \centering
    \includegraphics[width=8.6cm]{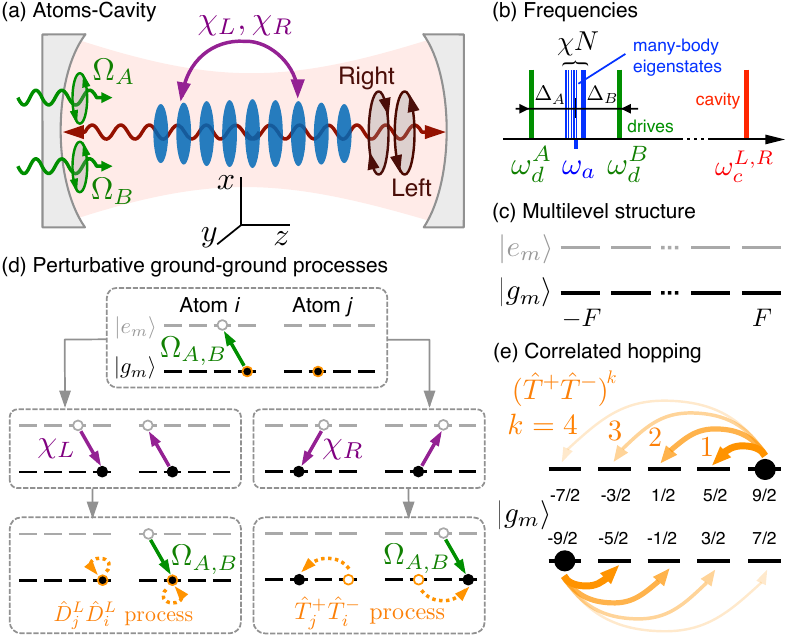}
    \caption{Effective ground state dynamics in a multilevel cavity QED system. (a) Sketch of the atom-cavity system. An ensemble of ultracold atoms (blue pancakes) are trapped in an optical cavity with quantization axis ($\hat{z}$) along the cavity axis. The cavity photons (red curve) with left-handed ($\sigma^{-}$) and right-handed ($\sigma^{+}$) circular polarizations mediate collective interactions ($\chi_{L,R}$) among the atoms. Two external drives are applied to the $\sigma^{-}$ polarized mode with atomic Rabi frequency $\Omega_{A,B}$. (b) Frequencies of cavity resonance, two external drives, and atomic transition dressed by photon-mediated interactions. (c) Multilevel internal level structure of the atoms, including ground ($|g_m\rangle$) and excited ($|e_m\rangle$) manifold. (d) Leading order interaction processes between atom $i$ and atom $j$ in the ground manifold. (e) Synthetic spatial ladder spanned by atomic ground manifold. The $\hat{T}^{+}\hat{T}^{-}$ processes can be understood as correlated hopping along each of the legs of the synthetic ladder.}
    \label{fig:overview}
\end{figure}

{\it System.}---We consider an ensemble of $N$ multilevel atoms confined in an optical cavity with quantization axis ($\hat{z}$) along the cavity axis, as depicted in Fig.~\ref{fig:overview}(a-c).
The internal level structure of each atom consists of a ground and an excited manifold with hyperfine spin $F_g$ and $F_e$ respectively.
The excited state manifold decays with spontaneous emission rate $\gamma$.
We consider a weak magnetic field limit such that all atomic transition frequencies can be approximated as a single $\omega_a$ \footnote{See Supplemental Material at [URL will be inserted by publisher] for details of adiabatic elimination, undepleted pump approximation, numerical results for dynamical phase transition and experimental parameters, includes Ref.~\cite{orioli2022emergent,muniz2020exploring,Boyd2007}.}.
Specifically, we label the ground states as $|g_m\rangle_i$, and the excited states as $|e_m\rangle_i$, where $m$ is the magnetic quantum number, and $i$ labels the atoms.
Two degenerate cavity modes with left-handed ($\sigma^{-}$) and right-handed ($\sigma^{+}$) circular polarization at frequency $\omega_c\equiv \omega_c^{L,R}$ couple to the transition between ground and excited manifolds with coupling strength $g_c$ and detuning $\Delta_c=\omega_c-\omega_a$. 
Two external $\sigma^{-}$ polarized lasers with frequencies $\omega_{d}^{A,B}$, detuned from the atomic transition by $\Delta_{A,B}=\omega_{d}^{A,B}-\omega_a$, are used to drive the cavity with intracavity Rabi frequency $\Omega_{A,B}$ respectively.

We focus on the far-detuned regime of the cavity ($|\Delta_c|\;\gg |\Delta_{A,B}|$), where the frequencies of laser drives are closer to the atomic transition rather than the cavity resonance.
In this regime, one can first adiabatically eliminate the injected light fields and intracavity fluctuations assuming $|\Delta_c|\;\gg g_c\sqrt{N},\kappa$, with $\kappa$ the cavity intensity decay rate.
The system is thus well described by an atom-only Hamiltonian with photon-mediated elastic interactions \cite{orioli2022emergent,Note1},
\begin{equation}
    \begin{aligned}
    \hat{H}/\hbar&=\omega_a\hat{N}_e+\chi_{L}\hat{L}^{+}\hat{L}^{-}+\chi_{R}\hat{R}^{+}\hat{R}^{-}\\
    &+\left[\left(\Omega_Ae^{-i\omega_d^{A}t}+\Omega_Be^{-i\omega_d^{B}t}\right)\hat{L}^{+}+h.c.\right],\\
    \end{aligned}
    \label{eq:hamil}
\end{equation}
where $\chi\equiv\chi_{L,R}=-g_c^2/\Delta_c$ is the photon-mediated interaction strength.
Here, $\hat{N}_e=\sum_{im}|e_m\rangle\langle e_m|_i$ is the atom number operator for the excited manifold, $\hat{L}^{+}=\sum_{im}C_m^{-1}|e_{m-1}\rangle\langle g_m|_i$, $\hat{L}^{-}=(\hat{L}^{+})^{\dag}$ are multilevel dipole operators with $\sigma^{-}$ polarization, and $\hat{R}^{+}=\sum_{im}C_m^{+1}|e_{m+1}\rangle\langle g_m|_i$,  $\hat{R}^{-}=(\hat{R}^{+})^{\dag}$ are multilevel dipole operators with $\sigma^{+}$ polarization, where $C^{p}_{m}\equiv \langle F_g,m;1,p|F_e,m+p\rangle$ are the Clebsch-Gordan coefficients.

The photon-mediated interactions in $\hat{H}$ [Eq.~(\ref{eq:hamil})] exchange excitations among atoms and generate a rich many-body spectrum of collective states [see Fig.~\ref{fig:overview}(b)], assuming $|\chi N|\;\gg \gamma$.
In this regime with weak driving fields ($|\Delta_{A,B}|,|\chi N|\;\gg |\Omega_{A,B}|$), as shown in \cite{Note1}, the many-body excited states are only virtually populated and can be adiabatically eliminated, giving rise to net interactions in the atomic ground manifold described by the following effective ground-state Hamiltonian,
\begin{equation}
    \hat{H}_{\mathrm{eff}}/\hbar=\sum_{\nu=A,B}\frac{|\Omega_{\nu}|^2\Delta_{\nu}}{\chi}\bigg[\Delta_{\nu}-\chi \hat{D}_L-\chi^2 \hat{T}^{+}\hat{G}_R^{\nu}\hat{T}^{-}\bigg]^{-1},
    \label{eq:heff}
\end{equation}
where $\hat{G}_R^{\nu}=(\Delta_{\nu}-\chi \hat{D}_R)^{-1}$. 
Here, $\hat{D}_L=\hat{P}_g\hat{L}^{-}\hat{L}^{+}\hat{P}_g$, $\hat{D}_R=\hat{P}_g\hat{R}^{-}\hat{R}^{+}\hat{P}_g$, $\hat{T}^{+}=\hat{P}_g\hat{L}^{-}\hat{R}^{+}\hat{P}_g$, and $\hat{T}^{-}=(\hat{T}^{+})^{\dag}$ are operators acting only on the ground manifold as ensured by $\hat{P}_g$, which is defined as the projection operator of the atomic ground states. 
In the case of $F_g=F_e=F$ as we explore in this paper, these operators can be expressed as $\hat{D}_{L,R}=\sum_i \hat{D}_i^{L,R}/\mathcal{N}_F$ and $\hat{T}^{+}=\sum_i\hat{T}^{+}_i/\mathcal{N}_F$, where
\begin{equation}
    \hat{D}^L_i=\hat{S}_i^{+}\hat{S}_i^{-}, \quad \hat{D}^R_i=\hat{S}_i^{-}\hat{S}_i^{+}, \quad
    \hat{T}^{+}_i=-\hat{S}_i^{+}\hat{S}_i^{+},
    \label{eq:opera}
\end{equation}
$\mathcal{N}_F=2F(F+1)$ is the normalization factor and $\hat{S}_i^{\pm}$ are raising and lowering spin-$F$ operators acting on atom $i$.
Note that $\hat{H}_{\mathrm{eff}}$ is fully collective and thus couples an atom $i$ with any other atom $j$ in the ensemble.
If the atoms start in the permutationally symmetric manifold, or on a direct product state of permutationally symmetric subsystems, they will remain there, and the scaling of Hilbert space dimension with atom number $N$ is reduced from exponential to polynomial.
Thanks to this simplification, we perform all the numerical calculations using Eq.~(\ref{eq:heff}).
However, the underlying physics in $\hat{H}_{\mathrm{eff}}$, which includes a sum over multi-body interactions, is still extremely complex even in these restricted Hilbert space. 

To gain physical intuition into Eq.~(\ref{eq:heff}), one can expand $\hat{H}_{\mathrm{eff}}$ in a power series of $\chi N/\Delta_{A,B}$, and keep only the leading order terms. This should be a good approximation in the off-resonant regime ($|\Delta_{A,B}|\;\gtrsim |\chi N|$). The effective Hamiltonian simplifies to 
\begin{equation}
    \hat{H}_{\mathrm{eff}}/\hbar\approx\sum_{\nu=A,B}\frac{|\Omega_{\nu}|^2}{\Delta_{\nu}}\hat{D}_L+\frac{|\Omega_{\nu}|^2\chi}{\Delta_{\nu}^2}(\hat{D}_L\hat{D}_L+\hat{T}^{+}\hat{T}^{-}).
    \label{eq:happrx}
\end{equation}
The first term in Eq.~(\ref{eq:happrx}) describes single-particle AC Stark shifts generated by each light field injected into the cavity, and the second term describes the leading order interactions in the ground manifold [see Fig.~\ref{fig:overview}(d)]. 
It consists of two terms:
The $\hat{D}_L\hat{D}_L$ term is diagonal and generates population-dependent collective shifts on the ground state levels which can give rise to spin squeezing \cite{Pezze2018}, while the $\hat{T}^{+}\hat{T}^{-}$ term generates transitions between levels.
Explicitly, it describes the processes where atom $i$ moves two internal levels down ($\hat{T}_i^{-}$) while atom $j$ moves two levels up ($\hat{T}_j^{+}$). 
Assuming hyperfine spin $F$ is a half-integer, it is convenient to visualize the atomic ground manifold as a synthetic two-leg ladder, where the upper and lower legs are the set of internal states directly connected by the $\hat{T}^{\pm}$ operators [see Fig.~\ref{fig:overview}(e)]. Under this concept, the $\hat{T}^{+}\hat{T}^{-}$ term is equivalent to correlated hopping along the legs of the synthetic ladder which generates strong correlations despite having no direct hopping processes between the legs. Note that the $\hat{T}^{+}$ and $\hat{T}^{-}$ hops can happen in different legs or both within the same leg. 
In order to better understand the Hamiltonian dynamics, it is useful to write the operators acting only on the ground manifold in terms of Schwinger bosons \cite{Note1},
\begin{equation}
    \begin{gathered}
    \hat{D}_L=\sum_{m}(C_{m}^{-1})^2\hat{a}_m^{\dag}\hat{a}_m, \quad \hat{D}_R=\sum_{m}(C_{m}^{+1})^2\hat{a}_m^{\dag}\hat{a}_m,\\
    \hat{T}^{+}=\sum_{m}C_{m}^{+1}C_{m+2}^{-1}\hat{a}_{m+2}^{\dag}\hat{a}_m,
    \end{gathered}
    \label{eq:schwinger}
\end{equation}
where $\hat{a}_m$ is the bosonic annihilation operator for state $|g_m\rangle$.
For simplicity, we also set the strength of the external drives to $\Omega\equiv\Omega_{A,B}=0.05\chi N$, and analyze the unitary dynamics by varying the detunings $\Delta_{A,B}$ of the laser drives.
By appropriate choices of $\Delta_{A,B}$, we can suppress the single-particle and collective shifts at $t=0$, and make the correlated hopping terms as the dominant process in the synthetic ladder \cite{Note1}.
 
\begin{figure}[t]
    \centering
    \includegraphics[width=8.6cm]{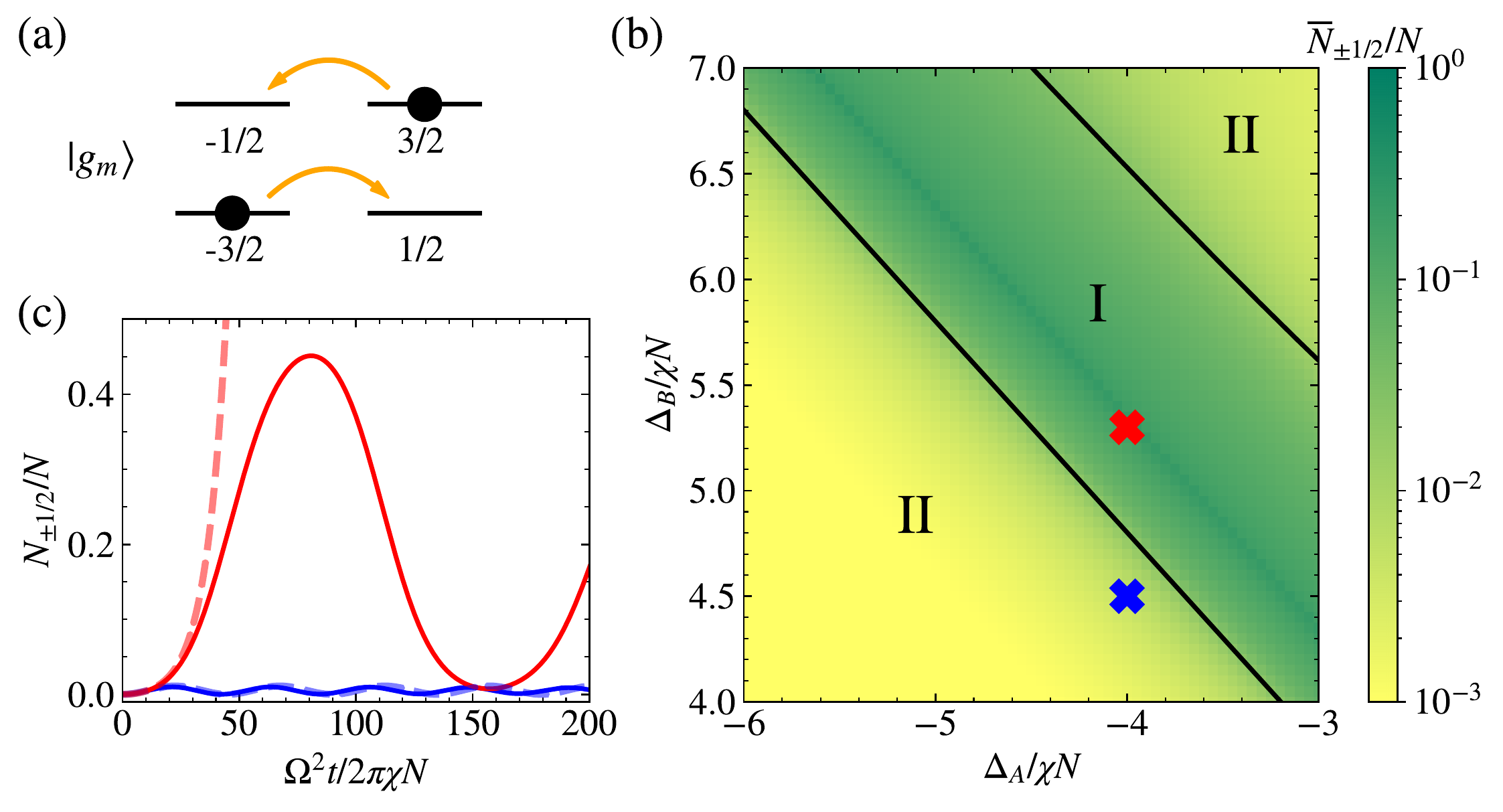}
    \caption{Pair production dynamics in a synthetic 4-level ladder. (a) The black dots show the initial state of the atoms and  the orange arrows show the correlated hopping process. (b) Dynamical phase diagram of pair production process. The black lines are the phase boundary separating phase I (with pair production) and phase II (without pair production). (c) Short-time dynamics in phase I (red curves) and phase II (blue curves). The choices of detunings $\Delta_{A,B}$ are indicated by cross marks in (b) with corresponding color. The solid lines are calculated by exact diagnoalization (ED), while the dashed lines are based on undepleted pump approximation (UPA).}
    \label{fig:pair}
\end{figure}

{\it Pair production.}---One of the simplest cases to understand the correlated hopping process ($\hat{T}^{+}\hat{T}^{-}$) is a system with a 4-level $F=3/2$ ground state manifold [see Fig.~\ref{fig:pair}(a)]. 
Considering the initial state $|g_{-3/2}\rangle^{\otimes(N/2)}|g_{3/2}\rangle^{\otimes(N/2)}$, the role of the $\hat{T}^{+}\hat{T}^{-}$ term is to generate correlated atom pairs in the initially unoccupied states $|g_{-1/2}\rangle$ and $|g_{1/2}\rangle$. 
We show that $\hat{H}_{\mathrm{eff}}$ [Eq.~(\ref{eq:heff})] in this system features an abrupt change of dynamical behavior as we tune the system parameters, i.e. dynamical phase transition (DPT). 
This type of DPT generated by pair production processes can be analyzed at both short times and long times \cite{Yang2019}. 

The short-time dynamics can be understood via undepleted pump approximation (UPA), where to the leading order one can replace the bosonic operators for macroscopically occupied states as c-numbers, $\hat{a}_{\pm 3/2}, \hat{a}^{\dag}_{\pm 3/2}\sim \sqrt{N/2}$. 
Under UPA, $\hat{H}_{\mathrm{eff}}$ [Eq.~(\ref{eq:heff})] becomes quadratic and therefore can be diagonalized analytically:
\begin{equation}
    \begin{aligned}
    \hat{H}_{\mathrm{eff}}/\hbar&\approx K_1\hat{a}^{\dag}_{-1/2}\hat{a}_{-1/2}+K_2\hat{a}^{\dag}_{1/2}\hat{a}_{1/2}\\
    &+K_3(\hat{a}_{-1/2}\hat{a}_{1/2}+h.c.).
    \label{eq:upa}
    \end{aligned}
\end{equation}
Here, $K_1$, $K_2$ and $K_3$ can be expressed as functions of $\Delta_{A,B}/\chi N$ \cite{Note1}.
The term proportional to $K_3$ is responsible for generating correlated atom pairs, while the terms proportional to $K_{1,2}$ impose an energy penalty for the pair production. 
Note that Eq.~(\ref{eq:upa}) is equivalent to the two-mode squeezing Hamiltonian well known in quantum optics \cite{scully1997quantum} and spinor BEC systems \cite{Pezze2018}. 
At short times when UPA is valid, as shown in Fig.~\ref{fig:pair}(c), one observes exponential growth of atom population $N_{\pm 1/2}$ in the initially unoccupied states $|g_{\pm 1/2}\rangle$ (red curves) in phase I (with pair production), which is described by $(K_1+K_2)^2<4K_3^2$.
Instead in phase II (without pair production), described by $(K_1+K_2)^2>4K_3^2$, one observes small oscillations of atom population (blue curves). 
The dynamical critical points are located at $(K_1+K_2)^2=4K_3^2$ [see black lines in Fig.~\ref{fig:pair}(b)].

To analyze the dynamics generated by $\hat{H}_{\mathrm{eff}}$ [Eq.~(\ref{eq:heff})] at longer times, we use exact diagonalization (ED) with 100 atoms.
At long times the DPT is signaled by a sharp change in behavior of the long-time average fractional population, $\overline{N}_{\pm 1/2}/N=\lim_{T\rightarrow\infty}\int_0^Tdt\, N_{\pm 1/2}(t)/(NT)$, which serves as an order parameter and distinguishes the two dynamical phases [see Fig.~\ref{fig:pair}(b)].
Phase I is characterized by non-zero $\overline{N}_{\pm 1/2}/N$, while phase II is characterized by $\overline{N}_{\pm 1/2}/N\approx 0$.
We analyze the critical exponents of this DPT in \cite{Note1}.
Our discussions of the 4-level system can be generalized to larger synthetic ladders directly, and the detunings $\Delta_{A,B}$ serve as control knobs of the correlated hopping process.

\begin{figure}[t]
    \centering
    \includegraphics[width=8.6cm]{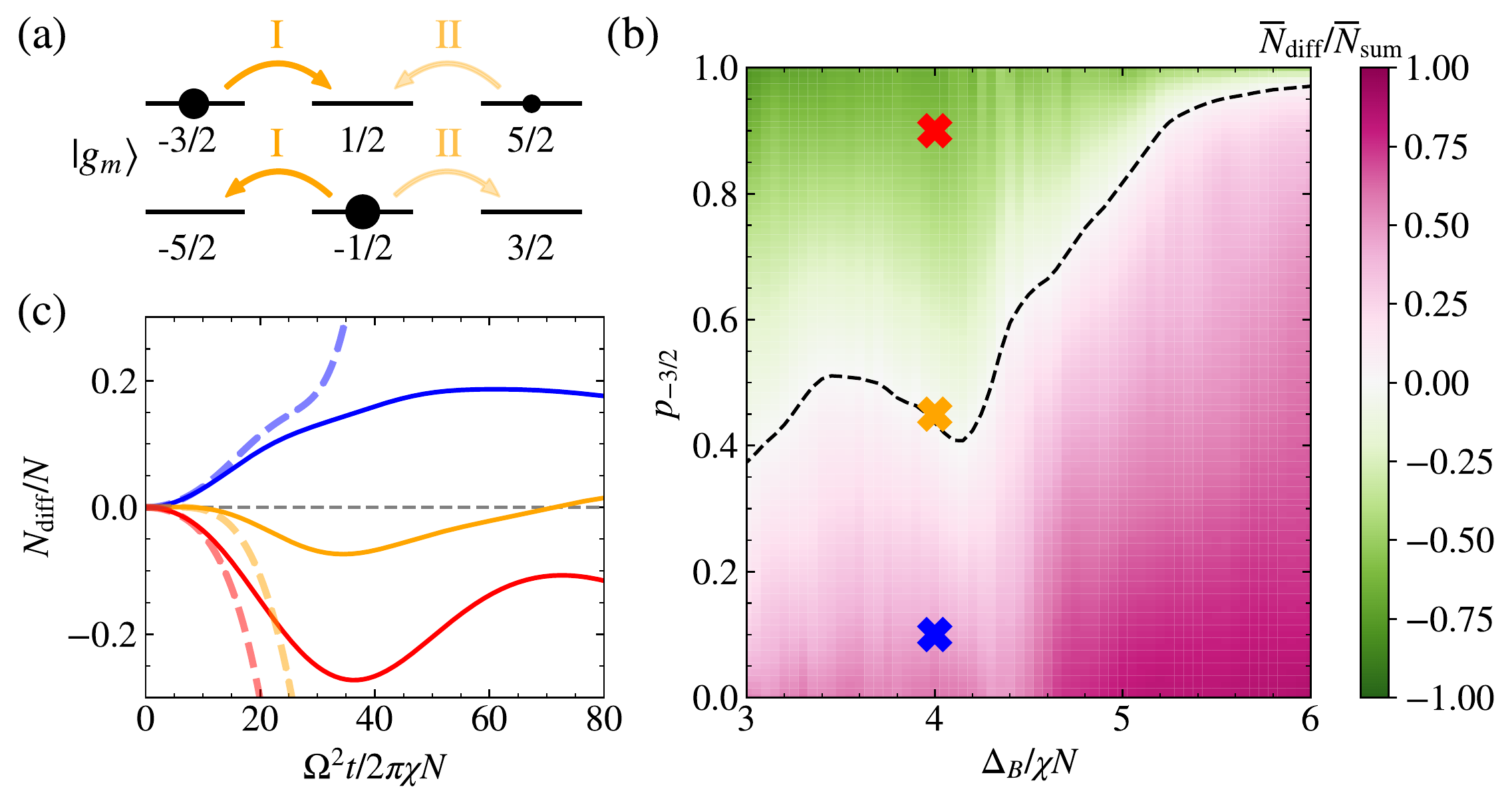}
    \caption{Chiral transport in a synthetic 6-level ladder. (a) The strength of the relevant correlated hopping processes (orange arrows) is indicated by the opacity of the arrows, which depends on the atom distribution in the upper leg (black dots). (b) Chiral transport behavior in the lower leg depends on the initial probability of occupying the state $|g_{-3/2}\rangle$ ($y$ axis) and the detuning of the external drive $B$ ($x$ axis) for fixed drive $A$ detuning which is set to $\Delta_A=-3\chi N$. The dashed line indicates balanced transport. (c) Short-time dynamics of population imbalance with parameters indicated by cross marks in (b) using the same color. The solid lines are calculated by ED, while the dashed lines are based on UPA.}
    \label{fig:chiral}
\end{figure}

{\it Chiral transport.}---
Internal atomic structure with larger number of levels allows for the observation of richer dynamical behaviors, including interaction-induced chiral transport. 
As shown in Fig.~\ref{fig:chiral}(a), we map a 6-level $F=5/2$ ground state manifold as a synthetic ladder, and initialize atoms to the center of the lower leg ($|g_{-1/2}\rangle$).
The chiral transport in the lower leg can be characterized by the population difference between atoms hopping to the right side ($|g_{3/2}\rangle$) and the left side ($|g_{-5/2}\rangle$), $N_{\mathrm{diff}}=N_{3/2}-N_{-5/2}$.
The chiral transport is not a consequence of the external drive polarization.
Suppose there are no atoms in the upper leg, the only relevant correlated hopping process will generate atom pairs in the state $|g_{-5/2}\rangle$ and $|g_{3/2}\rangle$, which leads to $N_{\mathrm{diff}}=0$.
Nevertheless, putting atoms in the upper chain gives rise to extra correlated hopping processes, in which the process I and II shown in Fig.~\ref{fig:chiral}(a) become the dominant processes and break left-right symmetry at short time \cite{Note1}.
If process I is stronger than process II, we have chiral transport to the left side of the lower chain ($N_{\mathrm{diff}}<0$), and vice versa.

We analyze the chiral transport behavior via ED of $\hat{H}_{\mathrm{eff}}$ [Eq.~(\ref{eq:heff})] with 20 atoms.
The initial state of this calculation is $[\sqrt{p_{-3/2}}|g_{-3/2}\rangle+\sqrt{1-p_{-3/2}}|g_{5/2}\rangle]^{\otimes (N/2)}|g_{-1/2}\rangle^{\otimes (N/2)}$, where $p_{-3/2}$ is the initial probability of occupying the state $|g_{-3/2}\rangle$ for the atoms in the upper leg.
The normalized longtime average $\overline{N}_{\mathrm{diff}}/\overline{N}_{\mathrm{sum}}$ as a function of  $p_{-3/2}$ and $\Delta_B$, where $N_{\mathrm{sum}}=N_{3/2}+N_{-5/2}$, is shown in Fig.~\ref{fig:chiral}(b) for fixed $\Delta_A$.   
It can be seen that for different choices of $\Delta_B$ it is possible to turn on both processes (I and II) if $\Delta_B/\chi N\in (3,4)$, or mainly turn on process II if $\Delta_B/\chi N\in (5,6)$.
Enforcing balanced transport [see the dashed line in Fig.~\ref{fig:chiral}(b)] requires equal weight of process I and II in the former case, or suppression of both processes in the latter case.
In Fig.~\ref{fig:chiral}(c), we compare the short-time dynamics of chiral transport at $\Delta_B=4\chi N$ with different choices of $p_{-3/2}$, indicating that the transport direction is fully tunable via initial state in this case. Unlike the 4-level system discussed above, UPA is not able to provide a qualitative description of chiral transport behavior at long times.

\begin{figure}[t]
    \centering
    \includegraphics[width=8.6cm]{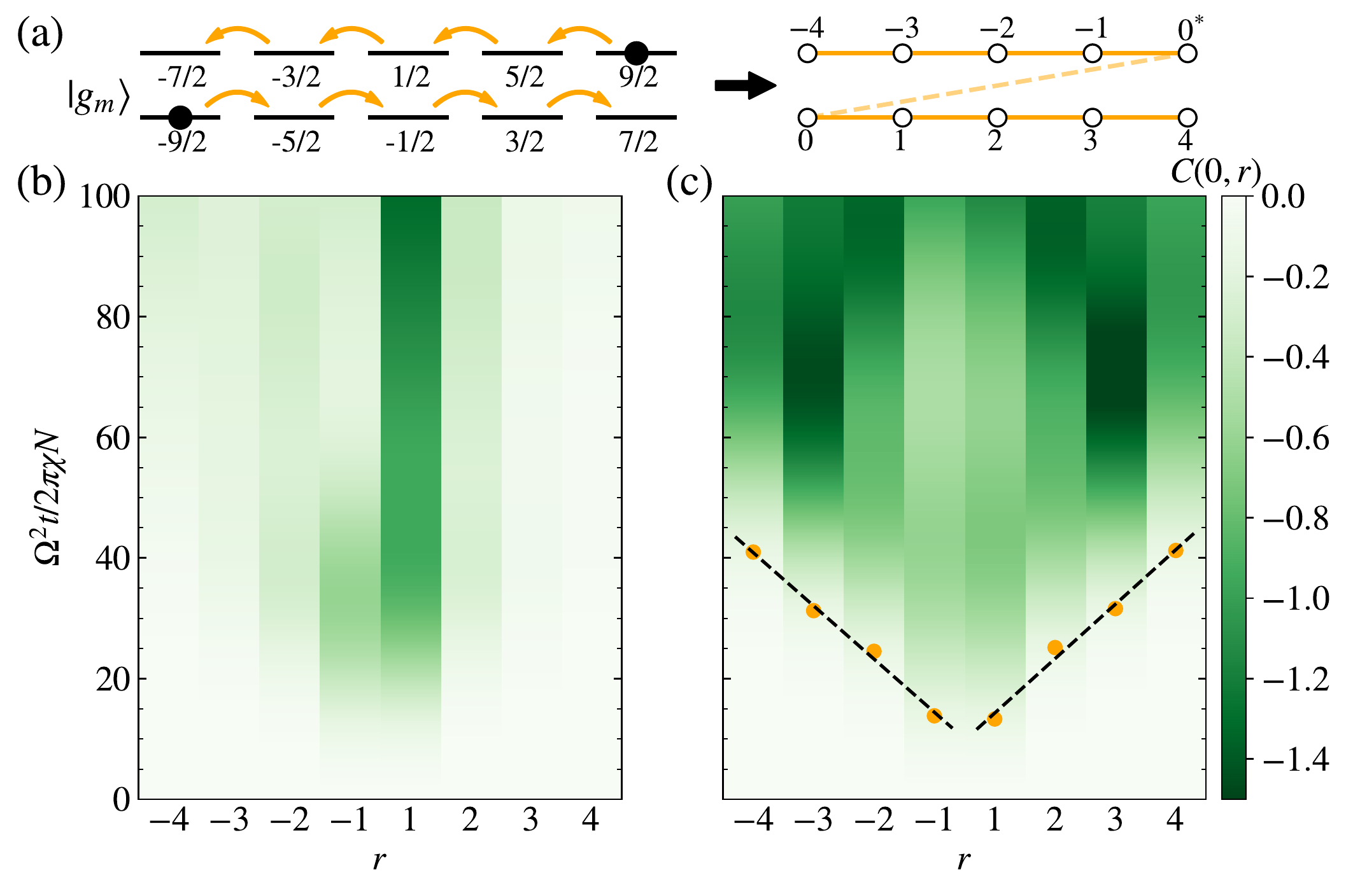}
    \caption{Correlation spreading in a synthetic 10-level ladder. (a) The  correlated hopping processes (orange arrows) for the initial state (black dots) allow us to assign position indices based on the direction of hopping. The orange dashed line indicates the spread of correlations between legs without direct hopping processes. (b) The atom number two-point correlations (see text) are  restricted to  nearest-neighbor sites for  $\Delta_A=-3\chi N$, $\Delta_B=3.7\chi N$, (c) but can undergo light-cone spreading for $\Delta_A=-3\chi N$, $\Delta_B=4.1\chi N$.}
    \label{fig:corr}
\end{figure}

{\it Correlation spreading.}--- Regardless of the all-to-all nature of the cavity mediated interactions, in our synthetic ladder we can engineer light-cone spreading of quantum correlations analogous to the one observed in real lattices with short-range or power-law interactions \cite{cheneau2012light,jurcevic2014quasiparticle,richerme2014non}. 
In Fig.~\ref{fig:corr}(a), we show a 10-level $F=9/2$ ground state manifold visualized as a synthetic ladder. For an initial product state $|g_{-9/2}\rangle^{\otimes(N/2)}|g_{9/2}\rangle^{\otimes(N/2)}$, it is possible to visualize the correlation spreading from a concatenated set of hopping processes [see orange arrows in Fig.~\ref{fig:corr}(a)] by assigning position indices for the synthetic lattice sites: $0$ and $0^{*}$ for the initial sites, and site labels  increasing to the  right and decreasing to the left.
Using this convention, we can analyze the correlation spreading in our synthetic  ladder, in terms of the two-point correlators $C(i,j)=\langle \hat{N}_i\hat{N}_j\rangle-\langle\hat{N}_i\rangle\langle\hat{N}_j\rangle$.

For the case of a system of 10 atoms, ED of $\hat{H}_{\mathrm{eff}}$ [Eq.~(\ref{eq:heff})] shows two distinct behaviors of $C(0,r)$ depending on the choice of $\Delta_B$ for fixed $\Delta_A$:
In one parameter regime, differential energy shifts imposed by $\Delta_B$ favor localization and the correlation is restricted to nearest-neighbor sites [see Fig.~\ref{fig:corr}(b)]; for another configuration, linear correlation spreading ($t\propto r$) to the whole system is energetically allowed [see Fig.~\ref{fig:corr}(c)]. The spreading is signaled by the appearance of a symmetric light cone [orange points in Fig.~\ref{fig:corr}(c)] in the two-point correlators, which is set at the time when $C(0,r)$ reaches $-0.15$ ($\sim 10\%$ of the maximum value).

{\it Experimental implementation.}---Our protocol can be directly implemented using fermionic alkaline-earth atoms in a cavity. 
The main advantage of these atoms is their unique atomic structure which offers simple ground (${}^1S_0$) and long-lived excited state manifolds (e.g. ${}^3P_1$) where we can explicitly isolate a single $F_g\to F_e$ transition, such as $5/2\to 5/2$ for ${}^{173}$Yb and $9/2\to 9/2$  for ${}^{87}$Sr. 
Although it might be possible to engineer similar correlated hopping processes in the ground hyperfine levels of alkali atoms, the proposed implementation is less direct since in this system the condition $|\chi N|\,\gg\gamma$ requires to make $\chi N$ comparable to the excited hyperfine splitting and as a consequence it is necessary to sum over the set of all excited hyperfine levels. 
For the particular case of ${}^{87}$Sr as discussed in \cite{Note1}, under current experimental conditions it is possible to operate in a regime where $|\chi N|/\gamma>10^2$ using $2\times 10^5$ atoms, so the dissipation during the time scale of interests can be ignored.
Moreover, our protocol can be directly generalized to the case with inhomogeneous atom-cavity couplings, $g_i=g_c\eta_i$.
The effective ground state Hamiltonian [Eq.~(\ref{eq:heff})] still takes the same form if we redefine the operators in Eq.~(\ref{eq:opera}) as $\hat{D}^L_i=\eta_i^2\hat{S}_i^{+}\hat{S}_i^{-}$, $\hat{D}^R_i=\eta_i^2\hat{S}_i^{-}\hat{S}_i^{+}$, $\hat{T}^{+}_i=-\eta_i^2\hat{S}_i^{+}\hat{S}_i^{+}$ \cite{Note1}.

{\it Conclusion and outlook.}---In summary, we propose an approach to engineer correlated hopping processes using cavity mediated interactions in a synthetic ladder spanned by atomic ground states, and discuss ways to observe pair production processes, chiral currents and emergent light-cone spreading of correlations. 
Here we discuss some of the intriguing features, but the complexity of these phenomena admits further explorations both theoretically and experimentally. Moreover, these are only a few of the many opportunities our quantum simulator can open. 
For example, using an additional transverse magnetic field, our protocol opens a path to engineer dynamical gauge fields, since the hopping phase in correlated hopping processes can be dynamically adjusted by the presence of other particles.  
Furthermore, even though here we assume a dilute gas and ignore contact interactions between atoms, by trapping atoms in 3D optical lattices, it is possible to make superexchange interactions comparable to the correlated hopping strength, opening a path for designing complex many-body Hamiltonians that are likely to display fast scrambling of quantum information and chaotic quantum behaviors \cite{Swingle}. 

\begin{acknowledgments}
We thank Aaron Friedman and Tobias Bothwell for useful discussions. 
This work is supported by the AFOSR Grant No. FA9550-18-1-0319, by the DARPA (funded via ARO) Grant No. W911NF-16-1-0576, the ARO single investigator Grant No. W911NF-19-1-0210, the NSF PHY1820885, NSF JILA-PFC PHY-1734006 and NSF QLCI-2016244 grants, by the DOE Quantum Systems Accelerator (QSA) grant and by NIST.
\end{acknowledgments}

\bibliography{reference}

\end{document}


\title{Photon-mediated correlated hopping in a synthetic ladder: Supplemental Materials}
\author{Anjun Chu}
\affiliation{JILA, NIST and Department of Physics, University of Colorado, Boulder, Colorado 80309, USA}
\affiliation{Center for Theory of Quantum Matter, University of Colorado, Boulder, Colorado 80309, USA}
\author{Asier Pi\~{n}eiro Orioli}
\affiliation{JILA, NIST and Department of Physics, University of Colorado, Boulder, Colorado 80309, USA}
\affiliation{Center for Theory of Quantum Matter, University of Colorado, Boulder, Colorado 80309, USA}
\author{Diego Barberena}
\affiliation{JILA, NIST and Department of Physics, University of Colorado, Boulder, Colorado 80309, USA}
\affiliation{Center for Theory of Quantum Matter, University of Colorado, Boulder, Colorado 80309, USA}
\author{James K. Thompson}
\affiliation{JILA, NIST and Department of Physics, University of Colorado, Boulder, Colorado 80309, USA}
\author{Ana Maria Rey}
\affiliation{JILA, NIST and Department of Physics, University of Colorado, Boulder, Colorado 80309, USA}
\affiliation{Center for Theory of Quantum Matter, University of Colorado, Boulder, Colorado 80309, USA}
\date{\today}

\maketitle

\section{Theory model}
In this section we describe how to obtain Eq.~(2) in the main text beginning with the fundamental interaction between atoms and light in a cavity. We work with ground and excited state manifolds of $N$ atoms, which are assumed to have hyperfine spin $F_g$ and $F_e$, respectively and are split by an energy $\omega_a$. The ground and excited states of atom $i$ are defined with respect to a quantization axis that is parallel to the cavity axis, and we label them as $\ket{g_m}_i$ and $\ket{e_m}_i$, respectively, where $m$ is the magnetic quantum number. These atoms interact with two circularly polarized cavity modes at frequency $\omega_c$, with strength $g_i$ for atom $i$. These coupling constants are parameterized as $g_i=g_c\eta_i$, where $\eta_i$ are dimensionless numbers of order 1 that describe the inhomogeneity in the couplings. The Hamiltonian describing this interaction is then \cite{orioli2022emergent}
\begin{equation}
    \hat{H}_{\text{atom-light}}/\hbar=\omega_a\hat{N}_e+\omega_c(\hat{a}^\dagger_L\hat{a}_L+\hat{a}^\dagger_R\hat{a}_R)+g_c(\hat{a}_L\hat{L}^++\hat{a}_R\hat{R}^++h.c.),
\end{equation}
where $\hat{a}_L(\hat{a}_R)$ are boson operators that annihilate a $\sigma^-$ ($\sigma^+$) photon, $\hat{N}_e=\sum_{i,m}\ket{e_m}\bra{e_m}_i$ is the number of atoms in the excited state, $\hat{L}^+=\sum_{i,m}\eta_i C_m^{-1}\ket{e_{m-1}}\bra{g_m}_i$ and $\hat{R}^+=\sum_{i,m}\eta_i C_m^{+1}\ket{e_{m+1}}\bra{g_m}_i$ are multilevel collective dipole operators with $\sigma^-$ and $\sigma^+$ polarization, and $C_{m}^p=\braket{F_g,m;1,p|F_e,m+p}$ are Clebsch-Gordan coefficients. The interaction describes the absorption of a $\sigma^-$ ($\sigma^+$) cavity photon  by the atomic ensemble accompanied by an atomic excitation with $\sigma^-$ ($\sigma^+$) light, together with the inverse process.

\begin{figure*}[t]
    \centering
    \includegraphics[width=14cm]{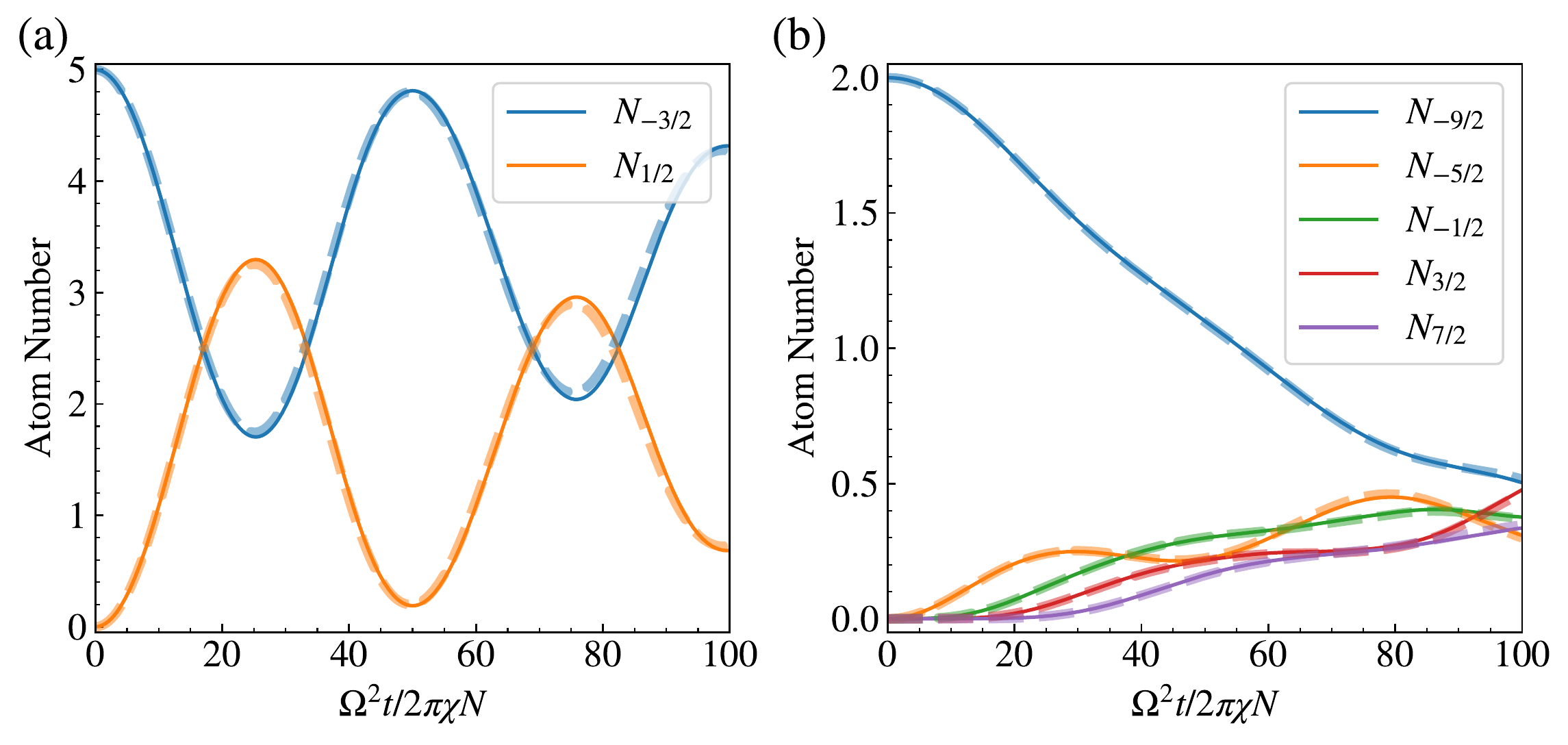}
    \caption{Numerical benchmarking of adiabatic elimination of atomic excited states. The Hamiltonian dynamics under Eq.~(\ref{eq:aad}) are shown in solid lines, while the Hamiltonian dynamics under Eq.~(\ref{eq:heff}) are shown in dashed lines. We set $\Delta_A=-3\chi N$, $\Delta_B=4.1\chi N$,
    $\Omega_{A,B}=\Omega=0.05\chi N$. We consider the following two cases: (a) 4-level system with initial state $|g_{-3/2}\rangle^{\otimes(N/2)}|g_{3/2}\rangle^{\otimes(N/2)}$, and atom number $N=10$. 
    (b) 10-level system with initial state $|g_{-9/2}\rangle^{\otimes(N/2)}|g_{9/2}\rangle^{\otimes(N/2)}$, and atom number $N=4$.}
    \label{fig:bench}
\end{figure*}

We consider the situation in which the cavity is far detuned from the atomic transition $|\Delta_c|\;\gg g_c\sqrt{N},\kappa$, where $\Delta_c=\omega_c-\omega_a$, and $\kappa$ is the cavity decay rate. In addition to this, the cavity will be driven by two tones with frequencies close to $\omega_a$, so they are far away from the cavity resonance frequency. Because of this, the number of intracavity photons is always very small, hence the light degrees of freedom can be adiabatically eliminated. The atom-light interaction, albeit weak, induces many-body splittings of the excited states, which the drives are designed to probe. The cavity-mediated atom-atom interactions are described by the following effective Hamiltonian \cite{orioli2022emergent}
\begin{equation}
    \hat{H}_{\text{atom-atom}}/\hbar=\omega_a\hat{N}_e+\chi\hat{L}^+\hat{L}^-+\chi\hat{R}^+\hat{R}^-,
\end{equation}
where $\hat{L}^-=(L^+)^\dagger$, $\hat{R}^-=(R^+)^\dagger$ and $\chi=-g_c^2/\Delta_c$ is the effective atom-atom interaction. The two injected driving fields ($A,B$) are assumed to be $\sigma^-$ polarized, with frequencies $\omega^{A,B}_d$ and intensities outside the cavity of $|\alpha_{A,B}|^2$ photons per second. These injected fields establish a small intracavity field that induces Rabi flopping on the atoms with Rabi frequencies $\Omega_{A,B}=-\sqrt{\kappa}\alpha_{A,B}g_c/\Delta_c$. Adding these terms leads to Eq.~(1) in the main text,
\begin{equation}
    \hat{H}_{\text{aa+d}}/\hbar=\omega_a\hat{N}_e+\chi\hat{L}^+\hat{L}^-+\chi\hat{R}^+\hat{R}^-+\Big[\Big(\Omega_A e^{-i\omega_d^A t}+\Omega_Be^{-i\omega_d^B t}\Big)\hat{L}^\dagger+h.c.\Big].
    \label{eq:aad}
\end{equation}
When the intracavity Rabi frequencies are small compare to ${\rm max}(\chi N, |\Delta_{A,B}|$), with $\Delta_{A,B}=\omega_d^{A,B}-\omega_a$, the excited state will only be virtually populated and can thus be adiabatically eliminated. To make the discussion simple, we consider first the case of a single drive $\Omega_A$, with associated Hamiltonian
\begin{equation}
    \hat{H}_{A}/\hbar=\omega_a\hat{N}_e+\chi\hat{L}^+\hat{L}^-+\chi\hat{R}^+\hat{R}^-+\Big(\Omega_A e^{-i\omega_d^A t}\hat{L}^++\Omega^{*}_A e^{i\omega_d^A t}\hat{L}^-\Big).
\end{equation}
In the rotating frame of the drive, this becomes
\begin{equation}
    \hat{H}_{A}'/\hbar=\underbrace{-\Delta_{A}\hat{N}_e+\chi\hat{L}^+\hat{L}^-+\chi\hat{R}^+\hat{R}^-}_{\hat{H}_0}+\underbrace{\Big(\Omega_A\hat{L}^++\Omega^{*}_A\hat{L}^-\Big)}_{\text{perturbation}}.
\end{equation}
Given that the Hamiltonian in this rotating frame is time-independent, the ground state effective description can be calculated using perturbation theory. First, notice that all the ground states are degenerate, with energy 0, in the absence of the perturbation. Furthermore, since the operator $\hat{L}^{+}$ maps the ground state manifold to single excitation states, the first order perturbative correction due to the drive is 0. The degeneracy of the ground states is lifted at second order using degenerate perturbation theory,
\begin{equation}
    \hat{H}_{\text{eff},A}/\hbar\approx -\hat{P}_g \Big(\Omega_A\hat{L}^++\Omega^{*}_A\hat{L}^-\Big)\frac{1}{\hat{H}_0}\Big(\Omega_A\hat{L}^++\Omega^{*}_A\hat{L}^-\Big)\hat{P}_g=-|\Omega_A|^2\hat{P}_g \hat{L}^-\frac{1}{\hat{H}_0}\hat{L}^+\hat{P}_g,
\end{equation}
where $\hat{P}_g$ are projectors onto the ground state manifold. The unperturbed Hamiltonian $\hat{H}_0$ includes not only the single particle term $\propto \hat{N}_e$ but also nonlinear contributions $\propto \hat{L}^+\hat{L}^-,\hat{R}^+\hat{R}^-$ and thus obtaining a simple expression for $\hat{H}_0^{-1}$ is not straightforward. Fortunately, $\hat{H}_0$ conserves the number of excitations, which is a strong enough symmetry to allow for an explicit calculation of $\hat{H}_0^{-1}$. To do this, consider the quantity
\begin{equation}
    \hat{H}_0\hat{L}^+\hat{P}_g=-\Delta_A\hat{N}_e\hat{L}^+\hat{P}_g+\chi \hat{L}^+\hat{L}^-\hat{L}^+\hat{P}_g+\chi \hat{R}^+\hat{R}^-\hat{L}^+\hat{P}_g.
\end{equation}
Due to the projectors, some parts of this expression can be simplified. First, $\hat{N}_e\hat{L}^+\hat{P}_g=\hat{L}^+\hat{P}_g$ since the image of the operator $\hat{L}^+\hat{P}_g$ is on the single excitation manifold. Second, the operators $\hat{L}^-\hat{L}^+\hat{P}_g$ and $\hat{R}^-\hat{L}^+\hat{P}_g$ preserve the ground state manifold and can thus be written as $\hat{L}^-\hat{L}^+\hat{P}_g\equiv\hat{D}_L\hat{P}_g$ and $\hat{R}^-\hat{L}^+\hat{P}_g\equiv\hat{T}^-\hat{P}_g$ where
\begin{equation}
    \begin{gathered}
        \hat{D}_L\equiv \hat{P}_g\hat{L}^{-}\hat{L}^{+}\hat{P}_g=\sum_{im}\eta_i^2(C_{m}^{-1})^2|g_m\rangle\langle g_m|_i,\\
        \hat{T}^{-}\equiv \hat{P}_g\hat{R}^{-}\hat{L}^{+}\hat{P}_g=\sum_{im}\eta_i^2C_{m-2}^{+1}C_{m}^{-1}|g_{m-2}\rangle\langle g_m|_i,\\
    \end{gathered}
    \label{eq:opera}
\end{equation}
are collective ground state operators only. Therefore

\begin{equation}\label{eqn:SuppSystem1}
   \boxed{ \hat{H}_0\hat{L}^+\hat{P}_g=\big(\hat{L}^+\hat{P}_g\big)\big(-\Delta_A+\chi\hat{D}_L\big)+\big(\hat{R}^+\hat{P}_g\big)\big(\chi\hat{T}^-\big)}.
\end{equation}
Note the ordering of the operators, which is important since $[\hat{L}^{\pm},\hat{D}_L]\neq 0$ and $[\hat{R}^{\pm},\hat{T}^-]\neq 0$. Similarly, 
\begin{equation}\label{eqn:SuppSystem2}
    \boxed{\hat{H}_0\hat{R}^+\hat{P}_g=\big(\hat{R}^+\hat{P}_g\big)\big(-\Delta_A+\chi\hat{D}_R\big)+\big(\hat{L}^+\hat{P}_g\big)\big(\chi\hat{T}^+\big)},
\end{equation}
where
\begin{equation}
    \begin{gathered}
         \hat{D}_R\equiv \hat{P}_g\hat{R}^{-}\hat{R}^{+}\hat{P}_g=\sum_{im}\eta_i^2(C_{m}^{+1})^2|g_m\rangle\langle g_m|_i,\\
        \hat{T}^{+}\equiv \hat{P}_g\hat{L}^{-}\hat{R}^{+}\hat{P}_g=\sum_{im}\eta_i^2C_{m}^{+1}C_{m+2}^{-1}|g_{m+2}\rangle\langle g_m|_i.
    \end{gathered}
    \label{eq:opera1}
\end{equation}
Right multiplying Eq.~(\ref{eqn:SuppSystem2}) by $(-\Delta_A+\chi\hat{D}_R)^{-1}\chi\hat{T}^-$ and subtracting the result from Eq.~(\ref{eqn:SuppSystem1}) results in
\begin{align}
    \begin{split}
        \hat{H}_0\Big[\hat{L}^+\hat{P}_g-\hat{R}^+\hat{P}_g\big(-\Delta_A+\chi\hat{D}_R\big)^{-1}\chi\hat{T}^-\Big]=\big(\hat{L}^+\hat{P}_g\big)\Big[\big(-\Delta_A+\chi\hat{D}_L\big)-\chi^2\hat{T}^+\big(-\Delta_A+\chi\hat{D}_R\big)^{-1}\hat{T}^-\Big].
    \end{split}
\end{align}
Left multiplying by $\hat{P}_g\hat{L}^-\hat{H}_0^{-1}$ leads to
\begin{align}
    \begin{split}
        \hat{D}_L-\hat{T}^+\big(-\Delta_A+\chi\hat{D}_R\big)^{-1}\chi\hat{T}^-=\big(\hat{P}_g\hat{L}^-\hat{H}_0^{-1}\hat{L}^+\hat{P}_g\big)\Big[\big(-\Delta_A+\chi\hat{D}_L\big)-\chi^2\hat{T}^+\big(-\Delta_A+\chi\hat{D}_R\big)^{-1}\hat{T}^-\Big].
    \end{split}
\end{align}
Therefore
\begin{equation}
    \big(\hat{P}_g\hat{L}^-\hat{H}_0^{-1}\hat{L}^+\hat{P}_g\big)=\frac{1}{\chi}-\frac{\Delta_A}{\chi}\bigg(\Delta_A-\chi\hat{D}_L-\chi^2\hat{T}^+\hat{G}_R^A\hat{T}^-\bigg)^{-1},
\end{equation}
where $\hat{G}_R^\nu=(\Delta_\nu-\chi\hat{D}_R)^{-1}$ and the effective ground state Hamiltonian due to drive $A$ is
\begin{equation}
    \boxed{\hat{H}_{\text{eff},A}/\hbar=-\frac{|\Omega_A|^2}{\chi}+\frac{|\Omega_A|^2\Delta_A}{\chi}\bigg(\Delta_A-\chi\hat{D}_L-\chi^2\hat{T}^+\hat{G}_R^A\hat{T}^-\bigg)^{-1}}.
\end{equation}
Similarly, drive $B$ will generate an analogous contribution to the effective ground state Hamiltonian. Since there are no possible interference pathways between the two drives (at second order) due to their different frequencies (we will take $\Delta_A\Delta_B<0$), the full effective Hamiltonian will be the sum of both contributions
\begin{equation}
    \boxed{\hat{H}_{\text{eff}}/\hbar=-\sum_{\nu=A,B}\frac{|\Omega_\nu|^2}{\chi}+\sum_{\nu=A,B}\frac{|\Omega_\nu|^2\Delta_\nu}{\chi}\bigg(\Delta_\nu-\chi\hat{D}_L-\chi^2\hat{T}^+\hat{G}_R^\nu\hat{T}^-\bigg)^{-1}}.
    \label{eq:heff}
\end{equation}
Omitting the c-number terms leads to Eq.~(2) in the main text. We have numerically benchmarked the validity of $\hat{H}_{\text{eff}}$ [Eq.~(\ref{eq:heff})] to capture  the dynamics of the full atomic Hamiltonian $\hat{H}_{\text{aa+d}}$ [Eq.~(\ref{eq:aad})]. Fig.~\ref{fig:bench} shows the excellent agreement between them.

Now we proceed to discuss the reason why we need two external drives. In the off-resonant regime ($|\Delta_{A,B}|\;\gtrsim |\chi N|$), one can expand $\hat{H}_{\mathrm{eff}}$ in a power series of $\chi N/\Delta_{A,B}$, and keep only the leading order terms as an excellent approximation. 
It's worth to mention that we only use this approximation for a qualitative understanding of the dynamics in our system, while all the quantitative calculations are still based on Eq.~(\ref{eq:heff}).
In this approximation, the effective Hamiltonian simplifies to 
\begin{equation}
    \hat{H}_{\mathrm{eff}}/\hbar\approx\sum_{\nu=A,B}\frac{|\Omega_{\nu}|^2}{\Delta_{\nu}}\hat{D}_L+\frac{|\Omega_{\nu}|^2\chi}{\Delta_{\nu}^2}(\hat{D}_L\hat{D}_L+\hat{T}^{+}\hat{T}^{-}).
\end{equation}
As we discussed in the main text, the first term describes the single-particle AC Stark shift, the second term describes the population dependent collective shift, and the third term describes correlated hopping. If we have only one external drive, the single-particle AC Stark shift term always play dominant role. In order to observe correlated hopping, one can use two external drives with detunings in opposite sign to suppress the undesirable energy shifts generated by the first two terms. Considering an initial state $|i\rangle$ and a final state $|f\rangle$ connected by correlated hopping processes, $\langle f|\hat{T}^{+}\hat{T}^{-}|i\rangle \neq 0$, the condition of suppressing the first two terms can be written as
\begin{equation}
    \bigg|\langle i|\hat{H}_{\mathrm{eff}}|i\rangle-\langle f|\hat{H}_{\mathrm{eff}}|f\rangle\bigg|\ll \bigg|\sum_{\nu=A,B}\frac{\hbar|\Omega_{\nu}|^2\chi}{\Delta_{\nu}^2}\langle f|\hat{T}^{+}\hat{T}^{-}|i\rangle\bigg|.
\end{equation}
This condition can be satisfied by tuning $\Delta_{A,B}$. Even though it is only computed using the initial state and thus only guarantee to be satisfied at $t=0$, we find that in a large parameter regime it allows to make the correlated hopping process to be dominant during a time window  where  they can give rise to non-trivial spreading through the synthetic ladder. 

\section{Schwinger bosons and undepleted pump approximation}
In the main text, we focused on the initial states as a direct product of a permutationally symmetric states of $N/2$ atoms on the upper leg of the synthetic ladder ($|\psi_{\mathrm{up}}\rangle$) and a permutationally symmetric state of $N/2$ atoms on the lower leg ($|\psi_{\mathrm{down}}\rangle$),
\begin{equation}
    |\psi\rangle=|\psi_{\mathrm{up}}\rangle\otimes|\psi_{\mathrm{down}}\rangle.
\end{equation}
Since the effective ground state Hamiltonian [Eq.~(\ref{eq:heff})] forbids any direct hopping process between the legs of the synthetic ladder, the Hamiltonian dynamics with these initial states occurs in a reduced Hilbert space consisted of any $SU\left(\frac{2F_g+1}{2}\right)\times SU\left(\frac{2F_g+1}{2}\right)$ rotations on $|\psi\rangle$, where $SU(n)$ denotes special unitary group of $n\times n$ matrices. So we can assign one set of Schwinger bosons $\{\hat{a}_{-F_g+1},\hat{a}_{-F_g+3},\cdots,\hat{a}_{F_g}\}$ to the upper leg and another set of Schwinger bosons $\{\hat{a}_{-F_g},\hat{a}_{-F_g+2},\cdots,\hat{a}_{F_g-1}\}$ to the lower leg. In this way, we can rewrite the operators acting only on the ground manifold as,
\begin{equation}
    \hat{D}_L=\sum_{m}(C_{m}^{-1})^2\hat{a}_m^{\dag}\hat{a}_m, \quad \hat{D}_R=\sum_{m}(C_{m}^{+1})^2\hat{a}_m^{\dag}\hat{a}_m,\quad
    \hat{T}^{+}=\sum_{m}C_{m}^{+1}C_{m+2}^{-1}\hat{a}_{m+2}^{\dag}\hat{a}_m,
    \label{eq:schwinger}
\end{equation}
where $\hat{a}_m$ is the bosonic annihilation operator for state $|g_m\rangle$. In the following, we will apply undepleted pump approximation (UPA) to the four-level and the six-level system discussed in the main text. The cascaded correlated hopping process in the ten-level system is beyond the reach of UPA.

\subsection{UPA for four-level system}
We first consider the initial state $|g_{-3/2}\rangle^{\otimes(N/2)}|g_{3/2}\rangle^{\otimes(N/2)}$ for the four-level ($F_g=F_e=3/2$) system discussed in the main text. The short time dynamics in this system can be understood via UPA, where to the leading order one can replace the bosonic operators of the initial  macroscopically occupied states as c-numbers, 
\begin{equation}
    \hat{a}_{\pm 3/2}= \hat{a}^{\dag}_{\pm 3/2}\approx \sqrt{\frac{N}{2}}, \quad {\rm and  }\quad   \hat{a}^{\dag}_{\pm 3/2}\hat{a}_{\pm 3/2}=\frac{N}{2}-\hat{a}^{\dag}_{\mp 1/2}\hat{a}_{\mp 1/2}.
\end{equation}
So the operators in Eq.~(\ref{eq:schwinger}) can be approximated as
\begin{equation}
    \hat{D}_L\approx \frac{N}{5}+\frac{8}{15}\hat{a}_{1/2}^{\dag}\hat{a}_{1/2}, \quad
    \hat{D}_R\approx \frac{N}{5}+\frac{8}{15}\hat{a}_{-1/2}^{\dag}\hat{a}_{-1/2}, \quad
    \hat{T}^{+}\approx -\frac{2\sqrt{6N}}{15}(\hat{a}_{1/2}^{\dag}+\hat{a}_{-1/2}).
\end{equation}

Plugging in this approximation in the effective ground state Hamiltonian [Eq.~(\ref{eq:heff})], and keeping the terms with operators $\hat{a}_{\pm 1/2}, \hat{a}_{\pm 1/2}^{\dag}$ up to second order, we obtain,
\begin{equation}
    \begin{aligned}
    \hat{H}_{\mathrm{eff}}/\hbar&=\sum_{\nu=A,B}\frac{|\Omega_{\nu}|^2\Delta_{\nu}}{\chi}\bigg[\Delta_{\nu}-\chi \hat{D}_L-\chi^2 \hat{T}^{+}(\Delta_{\nu}-\chi \hat{D}_R)^{-1}\hat{T}^{-}\bigg]^{-1}\\
    &\approx \sum_{\nu=A,B}\frac{|\Omega_{\nu}|^2\Delta_{\nu}}{\chi}\bigg[(\Delta_{\nu}-\chi N/5)-\frac{8\chi}{15}\hat{a}_{1/2}^{\dag}\hat{a}_{1/2}-\frac{\chi^2}{\Delta_{\nu}-\chi N/5}\frac{8N}{75}(\hat{a}_{1/2}^{\dag}+\hat{a}_{-1/2})(\hat{a}_{1/2}+\hat{a}_{-1/2}^{\dag})\bigg]^{-1}\\
    &\approx K_1\hat{a}^{\dag}_{-1/2}\hat{a}_{-1/2}+K_2\hat{a}^{\dag}_{1/2}\hat{a}_{1/2}+K_3(\hat{a}_{-1/2}^{\dag}\hat{a}_{1/2}^{\dag}+\hat{a}_{-1/2}\hat{a}_{1/2}).
    \end{aligned}
    \label{eq:upa1}
\end{equation}
where
\begin{equation}
    \begin{gathered}
        K_1=K_3=\sum_{\nu=A,B} \frac{8}{75}\frac{|\Omega_{\nu}|^2\Delta_{\nu}\chi N}{(\Delta_{\nu}-\chi N/5)^3},\\
        K_2=\sum_{\nu=A,B} \frac{8}{15}\frac{|\Omega_{\nu}|^2\Delta_{\nu}}{(\Delta_{\nu}-\chi N/5)^2}+\frac{8}{75}\frac{|\Omega_{\nu}|^2\Delta_{\nu}\chi N}{(\Delta_{\nu}-\chi N/5)^3}.
    \end{gathered}
\end{equation}

The effective Hamiltonian under the UPA [Eq.~(\ref{eq:upa1})] can be exactly diagonalized using a Bogoliubov transformation. In this case  the dynamics of the  populations in the $|g_{\pm 1/2}\rangle$ states, labelled as $N_{\pm 1/2}$, is found to be given by the following analytic formula in different parameter regimes:
\begin{equation}
    N_{\pm 1/2}(t)=\begin{cases}
        \displaystyle\frac{4K_3^2}{(K_1+K_2)^2-4K_3^2}\sin^2\bigg(\sqrt{(K_1+K_2)^2-4K_3^2}\,\frac{t}{2}\bigg) \quad\;\mathrm{if}\; (K_1+K_2)^2>4K_3^2\\
        \\
        \displaystyle \quad\quad\quad\quad\quad\quad\quad\quad\quad K_3^2t^2 \quad\quad\quad\quad\quad\quad\quad\quad\quad\quad\quad\quad\mathrm{if}\; (K_1+K_2)^2=4K_3^2\\
        \\
        \displaystyle\frac{4K_3^2}{4K_3^2-(K_1+K_2)^2}\sinh^2\bigg(\sqrt{4K_3^2-(K_1+K_2)^2}\,\frac{t}{2}\bigg) \;\;\;\mathrm{if}\; (K_1+K_2)^2<4K_3^2\\
    \end{cases},
\end{equation}
which are shown using dashed lines in Fig.~2(c) in the main text. Notice that $N_{\pm 1/2}(t)$ shows small sinusoidal oscillations in the regime $(K_1+K_2)^2>4K_3^2$, while it shows an exponential growth in the regime $(K_1+K_2)^2<4K_3^2$. This indicates a dynamical phase transition (DPT) at the critical point $(K_1+K_2)^2=4K_3^2$, which is equivalent to
\begin{equation}
    \sum_{\nu=A,B} \frac{|\Omega_{\nu}|^2\Delta_{\nu}}{(\Delta_{\nu}-\chi N/5)^2}=0 \quad \mathrm{or} \quad \sum_{\nu=A,B} \frac{|\Omega_{\nu}|^2\Delta_{\nu}}{(\Delta_{\nu}-\chi N/5)^2}+\frac{4}{5}\frac{|\Omega_{\nu}|^2\Delta_{\nu}\chi N}{(\Delta_{\nu}-\chi N/5)^3}=0.
\end{equation}
By solving these two equations, one can obtain the phase boundaries shown in Fig.~2(b) in the main text.

\subsection{UPA for six-level system}
In the main text, we consider the initial state $[\sqrt{p_{-3/2}}|g_{-3/2}\rangle+\sqrt{p_{5/2}}|g_{5/2}\rangle]^{\otimes (N/2)}|g_{-1/2}\rangle^{\otimes (N/2)}$ for the six-level ($F_g=F_e=5/2$) system, where $p_{5/2}=1-p_{-3/2}$. Similar to the previous subsection, we use the UPA to analyze the short time dynamics in this system. We have 6 independent Schwinger bosons in this problem,
\[\begin{aligned}
    &\mathrm{Empty:}\quad \hat{a}_{-5/2}, \quad \hat{a}_{3/2}, \quad \hat{a}_{1/2}, \quad \hat{a}_{-}=-\sqrt{p_{5/2}}\hat{a}_{-3/2}+\sqrt{p_{-3/2}}\hat{a}_{5/2}\\
    &\mathrm{Occupied:}\quad \hat{a}_{-1/2}, \quad \hat{a}_{+}=\sqrt{p_{-3/2}}\hat{a}_{-3/2}+\sqrt{p_{5/2}}\hat{a}_{5/2}\\
\end{aligned}\]
We would like to replace the occupied modes by c-numbers, and keep the empty modes up to second order. So we have
\begin{equation}
    \begin{aligned}
    \hat{D}_L&=\frac{16}{35}\hat{a}^{\dag}_{-1/2}\hat{a}_{-1/2}+\frac{16}{35}\hat{a}_{3/2}^{\dag}\hat{a}_{3/2}+\frac{18}{35}\hat{a}_{1/2}^{\dag}\hat{a}_{1/2}+\frac{2}{7}\hat{a}^{\dag}_{+}\hat{a}_{+}+\frac{2}{7}\hat{a}^{\dag}_{-}\hat{a}_{-}\\
    &\approx \frac{13}{35}N-\frac{16}{35}\hat{a}_{-5/2}^{\dag}\hat{a}_{-5/2}+\frac{8}{35}\hat{a}_{1/2}^{\dag}\hat{a}_{1/2},
    \end{aligned}
\end{equation}
\begin{equation}
    \begin{aligned}
    \hat{D}_R&=\frac{2}{7}\hat{a}_{-5/2}^{\dag}\hat{a}_{-5/2}+\frac{2}{7}\hat{a}_{3/2}^{\dag}\hat{a}_{3/2}+\frac{18}{35}\hat{a}^{\dag}_{-1/2}\hat{a}_{-1/2}+\frac{16}{35}\hat{a}_{1/2}^{\dag}\hat{a}_{1/2}+\frac{16}{35}(\sqrt{p_{-3/2}}\hat{a}^{\dag}_{+}-\sqrt{p_{5/2}}\hat{a}^{\dag}_{-})(\sqrt{p_{-3/2}}\hat{a}_{+}-\sqrt{p_{5/2}}\hat{a}_{-})\\
    &\approx \bigg(\frac{9}{35}+\frac{8}{35}p_{-3/2}\bigg)N-\frac{8}{35}\hat{a}_{-5/2}^{\dag}\hat{a}_{-5/2}-\frac{8}{35}\hat{a}_{3/2}^{\dag}\hat{a}_{3/2}+\frac{16}{35}(1-p_{-3/2})\hat{a}_{1/2}^{\dag}\hat{a}_{1/2}+\frac{16}{35}(p_{5/2}-p_{-3/2})\hat{a}_{-}^{\dag}\hat{a}_{-}\\
    &-\frac{16}{35}\sqrt{p_{-3/2}p_{5/2}}\sqrt{\frac{N}{2}}(\hat{a}^{\dag}_{-}+\hat{a}_{-}),
    \end{aligned}
\end{equation}
\begin{equation}
    \begin{aligned}
    \hat{T}^{+}&=-\bigg[\frac{4\sqrt{10}}{35}\hat{a}^{\dag}_{-1/2}\hat{a}_{-5/2}+\frac{12\sqrt{2}}{35}\hat{a}^{\dag}_{3/2}\hat{a}_{-1/2}+\frac{12\sqrt{2}}{35}\hat{a}^{\dag}_{1/2}(\sqrt{p_{-3/2}}\hat{a}_{+}-\sqrt{p_{5/2}}\hat{a}_{-})+\frac{4\sqrt{10}}{35}(\sqrt{p_{5/2}}\hat{a}^{\dag}_{+}+\sqrt{p_{-3/2}}\hat{a}^{\dag}_{-})\hat{a}_{1/2}\bigg]\\
    &\approx -\bigg[\frac{4\sqrt{10}}{35}\sqrt{\frac{N}{2}}\hat{a}_{-5/2}+\frac{12\sqrt{2}}{35}\sqrt{\frac{N}{2}}\hat{a}^{\dag}_{3/2}+\frac{12\sqrt{2}}{35}\sqrt{\frac{p_{-3/2}N}{2}}\hat{a}^{\dag}_{1/2}+\frac{4\sqrt{10}}{35}\sqrt{\frac{p_{5/2}N}{2}}\hat{a}_{1/2}\bigg].
    \end{aligned}
\end{equation}

In this way, the effective ground state Hamiltonian [see Eq.~(\ref{eq:heff})] can be approximated into the following form,
\begin{equation}
    \hat{H}_{\mathrm{eff}}/\hbar\approx\sum_{ij}\bigg[\mathcal{L}_{ij}\hat{a}_i^{\dag}\hat{a}_j+\frac{1}{2}\mathcal{M}_{ij}\hat{a}_i^{\dag}\hat{a}_j^{\dag}+\frac{1}{2}\mathcal{M}^{*}_{ij}\hat{a}_i\hat{a}_j\bigg],
    \label{eq:hbdg}
\end{equation}
where $i,j\in \{-5/2,1/2,3/2\}$, and $\hat{a}_{-}, \hat{a}^{\dag}_{-}$ are not included because they only appear in the terms beyond the second order. Here, $\mathcal{L}$ is a Hermitian matrix,  and $\mathcal{M}$ is a symmetric matrix. They take the following form:
\begin{equation}
    \mathcal{L}=\begin{pmatrix}
    K_1 & K_4\sqrt{p_{5/2}} & 0\\
    K_4\sqrt{p_{5/2}} & K_2 & K_5\sqrt{p_{-3/2}}\\
    0 & K_5\sqrt{p_{-3/2}} & K_3\\
    \end{pmatrix}, \quad 
    \mathcal{M}=\begin{pmatrix}
    0 & K_6\sqrt{p_{-3/2}} & K_6\\
    K_6\sqrt{p_{-3/2}} & 2K_6\sqrt{p_{-3/2}p_{5/2}} & K_6\sqrt{p_{5/2}}\\
    K_6 & K_6\sqrt{p_{5/2}} & 0\\
    \end{pmatrix},
\end{equation}
where
\begin{equation}
    \begin{gathered}
    K_1=-\frac{16}{35}C_1+\frac{16}{245}C_2, \quad K_2=\frac{8}{35}C_1+\bigg(\frac{144}{1225}p_{-3/2}+\frac{16}{245}p_{5/2}\bigg)C_2,\\
    K_3=K_5=\frac{144}{1225}C_2, \quad K_4=\frac{16}{245}C_2, \quad  K_6=\frac{48\sqrt{5}}{1225}C_2,\\
    C_1=\sum_{\nu=A,B}\frac{|\Omega_{\nu}|^2\Delta_{\nu}}{(\Delta_{\nu}-13\chi N/35)^2}, \quad C_2=\sum_{\nu=A,B}\frac{|\Omega_{\nu}|^2\Delta_{\nu}\chi N}{(\Delta_{\nu}-13\chi N/35)^2[\Delta_{\nu}-(9+8p_{-3/2}\chi N/35)]}.\\
    \end{gathered}
\end{equation}

\begin{figure*}[t]
    \centering
    \includegraphics[width=14cm]{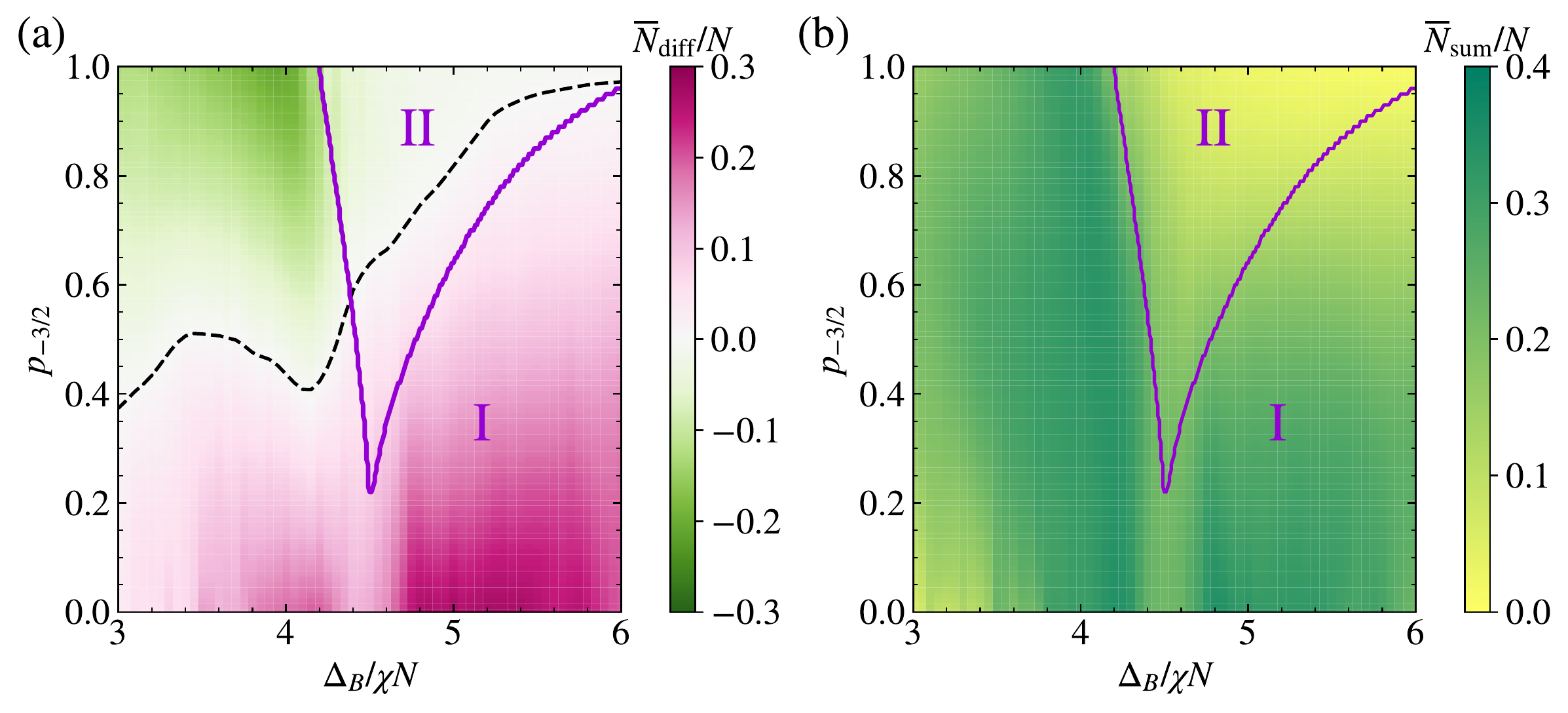}
    \caption{Dynamical phase boundary in the six-level system (purple line) calculated by UPA in comparison with (a) $\overline{N}_{\mathrm{diff}}/N$ and (b) $\overline{N}_{\mathrm{sum}}/N$ calculated by exact diagonalization with 20 atoms. We set $\Delta_A=-3\chi N$, and the same initial state for the six-level system as discussed in the main text. The black dashed line in (a) indicates balanced transport.}
    \label{fig:upa6}
\end{figure*}

The dynamics of an operator $\hat{O}$ under Eq.~(\ref{eq:hbdg}) can be solved exactly using the Heisenberg equations of motion $\partial_t \hat{O}=i[\hat{H},\hat{O}]/\hbar$, so we have the following set of equations for the bosonic creation and annilation operators, 
\begin{equation}
    \partial_t\begin{pmatrix}
    \hat{A}\\
    \hat{A}^{\dag}\\
    \end{pmatrix}=-i\begin{pmatrix}
    \mathcal{L} & \mathcal{M}\\
    -\mathcal{M}^{*} & -\mathcal{L}^{*}\\
    \end{pmatrix}\begin{pmatrix}
    \hat{A}\\
    \hat{A}^{\dag}\\
    \end{pmatrix},
    \label{eq:heisenbog}
\end{equation}
where $\hat{A}=(\hat{a}_{-5/2}, \hat{a}_{1/2}, \hat{a}_{3/2})^{T}$. This allows us to reduce the problem into the calculation of right eigenvalues and right eigenvectors for a non-Hermitian matrix,
\begin{equation}
    \begin{pmatrix}
    \mathcal{L} & \mathcal{M}\\
    -\mathcal{M}^{*} & -\mathcal{L}^{*}\\
    \end{pmatrix}\begin{pmatrix}
    \vec{u}_p\\
    \vec{v}_p\\
    \end{pmatrix}=\epsilon_p\begin{pmatrix}
    \vec{u}_p\\
    \vec{v}_p\\
    \end{pmatrix},
    \label{eq:bdg}
\end{equation}
These are the so-called Bogoliubov-de Gennes (BdG) equations. By defining  a matrix $\mathcal{T}$ whose column vectors are the right eigenvectors $(\vec{u}_p, \vec{v}_p)^{T}$, and a diagonal matrix $\mathcal{H}$ whose diagonal elements are the right eigenvalues $\epsilon_p$, we have
\begin{equation}
    \begin{pmatrix}
    \hat{A}(t)\\
    \hat{A}^{\dag}(t)\\
    \end{pmatrix}=\mathcal{U}_t\begin{pmatrix}
    \hat{A}(0)\\
    \hat{A}^{\dag}(0)\\
    \end{pmatrix}, \quad \mathcal{U}_t=\exp\left[-i\begin{pmatrix}
    \mathcal{L} & \mathcal{M}\\
    -\mathcal{M}^{*} & -\mathcal{L}^{*}\\
    \end{pmatrix}t\right]=\mathcal{T}e^{-i\mathcal{H}t}\mathcal{T}^{-1}.
    \label{eq:dyn6}
\end{equation}

Since we are considering the initial state as the vacuum state, the expectation values of any quadratic form of bosonic operators are given by
\begin{equation}
    \left\langle\mathrm{vac}\left|\begin{pmatrix}
    \hat{A}(t)\\
    \hat{A}^{\dag}(t)\\
    \end{pmatrix}\begin{pmatrix}
    \hat{A}(t) & \hat{A}^{\dag}(t)
    \end{pmatrix}\right|\mathrm{vac}\right\rangle=\mathcal{U}_t\begin{pmatrix}
    0 & \mathcal{I}\\
    0 & 0\\
    \end{pmatrix}\mathcal{U}_t^{T},
\end{equation}
where $\mathcal{I}$ is the identity matrix. Using this method, we can calculate the short time dynamics of $N_{\mathrm{diff}}$, which are shown with dashed lines in Fig.~3(c) in the main text.

Based on Eq.~(\ref{eq:dyn6}), one can separate between the exponential growth (Phase I) and the sinusoidal oscillation (Phase II): If all the right eigenvalues $\epsilon_p$ are real numbers, the system is in Phase II; If at least one of the right eigenvalues $\epsilon_p$ are complex numbers, the system is in Phase I. According to Eq.~(\ref{eq:bdg}), if $(\vec{u}_p, \vec{v}_p)^{T}$ is a right eigenvector with right eigenvalue $\epsilon_p$, $(\vec{v}_p^{*}, \vec{u}_p^{*})^{T}$ is a right eigenvector with right eigenvalue $-\epsilon_p^{*}$. Since all the element of matrix $\mathcal{L}$ and $\mathcal{M}$ are real numbers, if $\epsilon_p$ is a right eigenvalue, $\epsilon_p^{*}$ is also a right eigenvalue. Therefore, if $\epsilon_p$ is a real number, we have a pair of right eigenvalues $\epsilon_p,-\epsilon_p$; if $\epsilon_p$ is a complex number, we have four right eigenvalues $\epsilon_p,-\epsilon_p,\epsilon_p^{*},-\epsilon_p^{*}$. So we have two different cases for the six-level system discussed above: $2$ real eigenvalues, $4$ complex eigenvalues (Phase I); $6$ real eigenvalues (Phase II). The phase boundary calculated by UPA generally agrees with the long-time averages $\overline{N}_{\mathrm{diff}}=\overline{N}_{3/2}-\overline{N}_{-5/2}$ and $\overline{N}_{\mathrm{sum}}=\overline{N}_{3/2}+\overline{N}_{-5/2}$ calculated by exact diagonalization with 20 atoms [see Fig.~\ref{fig:upa6}].

We can now explain the short-time behavior of chiral transport discussed in the main text based on the $4$ pair creation processes described by matrix $\mathcal{M}$:
\begin{enumerate}[label=\arabic*)]
    \item Process I: $\hat{a}_{-5/2}^{\dag}\hat{a}_{1/2}^{\dag}$ with strength $K_6\sqrt{p_{-3/2}}$
    \item Process II: $\hat{a}_{3/2}^{\dag}\hat{a}_{1/2}^{\dag}$ with strength $K_6\sqrt{p_{5/2}}$
    \item Process III: $\hat{a}_{-5/2}^{\dag}\hat{a}_{3/2}^{\dag}$ with strength $K_6$
    \item Process IV: $\hat{a}_{1/2}^{\dag}\hat{a}_{1/2}^{\dag}$ with strength $K_6\sqrt{p_{-3/2}p_{5/2}}$
\end{enumerate}
Notice that only process I and II change the value of $N_{\mathrm{diff}}$, and the relative strength between these two processes can be tuned by $p_{-3/2}$, while $\Delta_B$ is another control knob to determine whether these processes are on resonance or not. When these two processes are both on resonance, e.g. $\Delta_B/\chi N\in (3,4)$, the direction of chiral transport is fully tunable via $p_{-3/2}$ [see Fig.~\ref{fig:upa6}(a)]. When process I is off resonance, e.g. $\Delta_B/\chi N\in (5,6)$, we always find chiral transport to the right side in the exponential growth regime, and balanced transport is only possible to achieved by tuning the whole system to sinusoidal oscillation regime [see Fig.~\ref{fig:upa6}(a)].

\section{Numerical results for dynamical phase transition} 
\begin{figure*}[t]
    \centering
    \includegraphics[width=17.8cm]{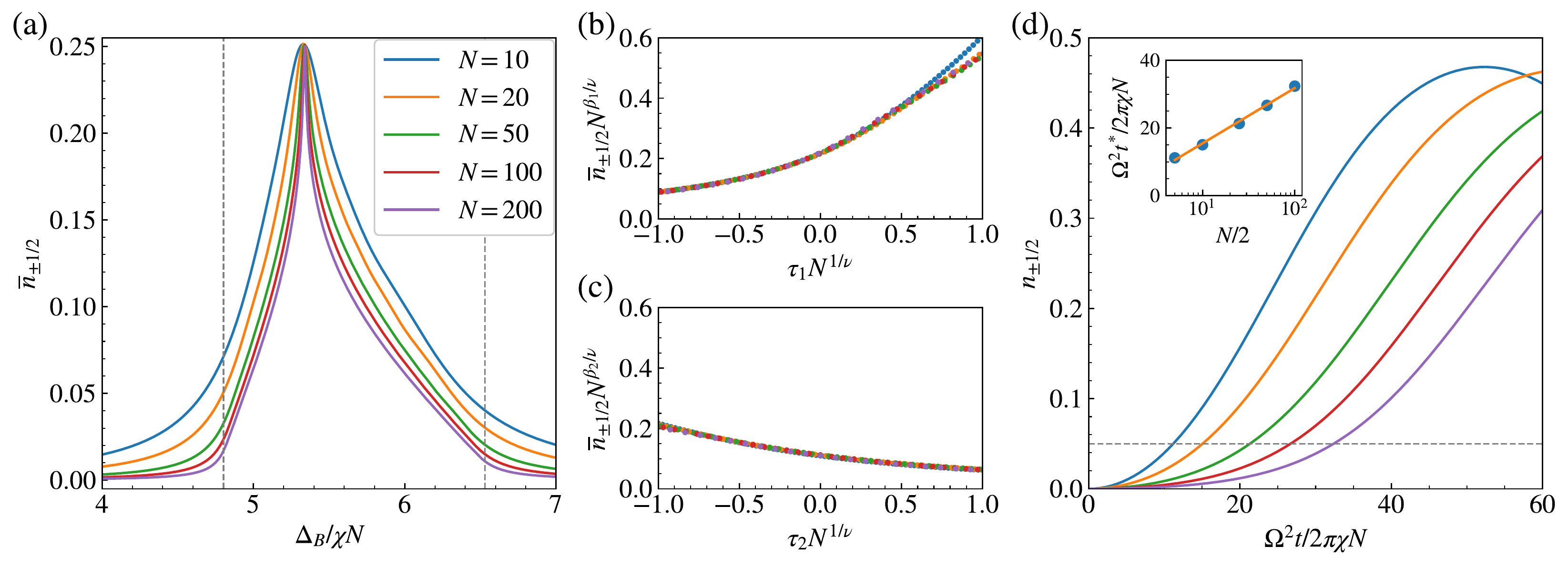}
    \caption{Finite size scaling of DPT by varying atom number $N$ while keeping $\chi N$ fixed. (a) The long-time average fractional atom population of $|g_{\pm 1/2}\rangle$ states ($\overline{n}_{\pm 1/2}$) at $\Delta_A=-4\chi N$. The dashed lines indicate the dynamical critical points predicted by UPA in the previous section. (b,c) Data collapse of finite-size calculations of $\overline{n}_{\pm 1/2}$ near the critical points $\Delta_{B,1}$ and $\Delta_{B,2}$ (see text). (d) Short-time dynamics of $n_{\pm 1/2}$ at $\Delta_A=-4\chi N$ and $\Delta_B=5.3\chi N$. The dashed line marks $n_{\pm 1/2}=0.05$, which determines the typical time scale $t^{*}$ for the obsevation of the DPT. The inset shows the logarithmic scaling of $t^{*}$.}
    \label{fig:fss}
\end{figure*}

Here we present additional numerical results using exact diagonalization that complements the discussion of the dynamical phase transition (DPT) discussed in Fig.~2 of the main text. 
Since exact diagonalization is only possible for a small atom number, we explore the properties of this DPT in the thermodynamic limit using finite size scaling. Given the collective nature of the cavity-mediated interactions, it is convenient to vary the atom number $N$ while keeping $\chi N$ fixed. As we described in the main text, we consider the initial state $|g_{-3/2}\rangle^{\otimes(N/2)}|g_{3/2}\rangle^{\otimes(N/2)}$, which is an eigenstate of this system when $\Delta_A,\Delta_B\rightarrow\infty$. We choose $\Omega_A=\Omega_B=\Omega=0.05\chi N$, and then perform a sudden quench to $\Delta_A=-4\chi N$ and $\Delta_B$ between $4\chi N$ and $7\chi N$. The order parameter of this DPT is the long-time average of the fractional atom population of the  $|g_{\pm 1/2}\rangle$ states,
\begin{equation}
    \overline{n}_{\pm 1/2}=\lim_{T\rightarrow\infty}\frac{1}{T}\int_0^{T}\frac{N_{\pm 1/2}(t)}{N}dt,
\end{equation}
shows a sharp change behavior as we varying $\Delta_B$ [see Fig.~\ref{fig:fss}(a)]. Near the dynamical critical points ($\Delta_{B,1}=4.80\chi N$, $\Delta_{B,2}=6.53\chi N$), it is convenient to assume $\overline{n}_{\pm 1/2}$ as a homogeneous function,
\begin{equation}
    \overline{n}_{\pm 1/2}=N^{-\beta_{1,2}/\nu}f(\tau_{1,2}N^{1/\nu}),
\end{equation}
where $\tau_{1}=(\Delta_B-\Delta_{B,1})/\chi N$, $\tau_{2}=(\Delta_B-\Delta_{B,2})/\chi N$. The data collapse of finite-size calculations are shown in Fig.~\ref{fig:fss}(b,c), which gives $\beta_1=1.15$, $\beta_2=1.04$, $\nu=2.38$. Based on the data collapse, one can conclude that 
\begin{equation}
    \begin{gathered}
       \overline{n}_{\pm 1/2}=0\;(\tau_{1}<0), \quad \overline{n}_{\pm 1/2}\propto \tau_{1}^{\beta_1} \;(\tau_{1}>0),\\
       \overline{n}_{\pm 1/2}\propto (-\tau_2)^{\beta_2}\;(\tau_{2}<0), \quad \overline{n}_{\pm 1/2}=0 \;(\tau_{2}>0),
    \end{gathered}
\end{equation}
at thermodynamic limit with $\tau_{1,2}\rightarrow 0$.

In experiments, it is easier to explore the DPT using short-time dynamics instead of long-time averages. In phase I, the exponential growth of $n_{\pm 1/2}$ is triggered by quantum fluctuations in the initial state, and therefore there is a delay time for $|g_{\pm 1/2}\rangle$ states to accumulate a macroscopic population [see Fig.~\ref{fig:fss}(d)]. The delay time increases as we increase the atom number $N$. A reasonable measure of the delay time would be the time $t^{*}$ for $n_{\pm 1/2}$ reaching 0.05 ($\sim 10\%$ of maximum value). As shown in the inset of Fig.~\ref{fig:fss}(d), we find $t^{*}\propto \ln(N/2)$.

\section{Experimental parameters}
Here we discuss the specific case of the ${}^1S_0\rightarrow {}^3P_1$ transition in ${}^{87}$Sr using the experimental parameters    described in Ref.~\cite{muniz2020exploring}. 
We show that current experimental parameters  lie in the  regime  where our  theory predictions are valid and the dynamics is dominated by unitary evolution.  The conditions  are summarized in Table~\ref{tab:sum}. 
In this system the typical atom-light coupling strength is $g_c\sim 2\pi\times 10$kHz, and the cavity intensity decay rate is $\kappa\sim 2\pi\times 100$kHz. 
Choosing atom number $N\sim 2\times10^5$, and cavity detuning $\Delta_c=2\pi\times 20$MHz, we ensure the adiabatic elimination of the cavity photons and dominant unitary dynamics compared to cavity decay.  
As we discussed in the main text, setting $\Omega_{A,B}=\Omega=0.05\chi N$, and $\Delta_{A,B}$ comparable or larger than $\chi N$ satisfy the requirement of adiabatic elimination of atomic excited states. 
Note that the correlated hopping processes occur at a rate $\Omega^2\chi N/\Delta_{A,B}^2$, while the dissipation process in atomic ground manifold due to spontaneous emission from excited manifold occur at a rate $\Omega^2\gamma/\Delta_{A,B}^2$.
Since for the above parameters $\chi N\sim 2\pi\times 1$MHz, is more than two orders of magnitude larger than the spontaneous emission rate $\gamma=2\pi\times 7.5$kHz of ${}^3P_1$ states, we can ignore dissipation during the time scale of interests.
As shown in Fig.~4 in the main text, the correlation spreading takes place at a time scale  $\Omega^2t/2\pi\chi N\sim 100$ for atom number $N=10$, and the corresponding delay time is $\Omega^2t^{*}/2\pi\chi N\sim 10$.
For $N\sim 2\times10^5$, the delay time is expected to be $7$ times longer, and the relevant experimental time scale to observe correlation spreading is $\Omega^2t/2\pi\chi N\sim 200$, which corresponds to $t\sim 80$ms.

\begin{table}[h!]
  \renewcommand\arraystretch{1.5}
  \centering
  \caption{Summary of approximations in theory model and the required parameter regimes (see text for details)}
    \begin{tabular}{C{8cm}|C{5cm}}
      \hline 
      Theory approximations & Required parameter regimes \\
      \hline 
      Adiabatic elimination of cavity photons  & $|\Delta_c|\;\gg g_c\sqrt{N}$ \\
      Negligible cavity loss  & $|\Delta_c|\;\gg \kappa$ \\
      Adiabatic elimination of atomic excited states & $|\Delta_{A,B}|,|\chi N|\;\gg |\Omega_{A,B}|$ \\
      Negligible spontaneous emission  & $|\chi N|\;\gg \gamma$ \\
      Weak magnetic field limit & $|\chi N|\;\gg \delta_e F_e$ \\
      \hline
    \end{tabular}
    \label{tab:sum}
\end{table}

\begin{figure*}[t]
    \centering
    \includegraphics[width=17.8cm]{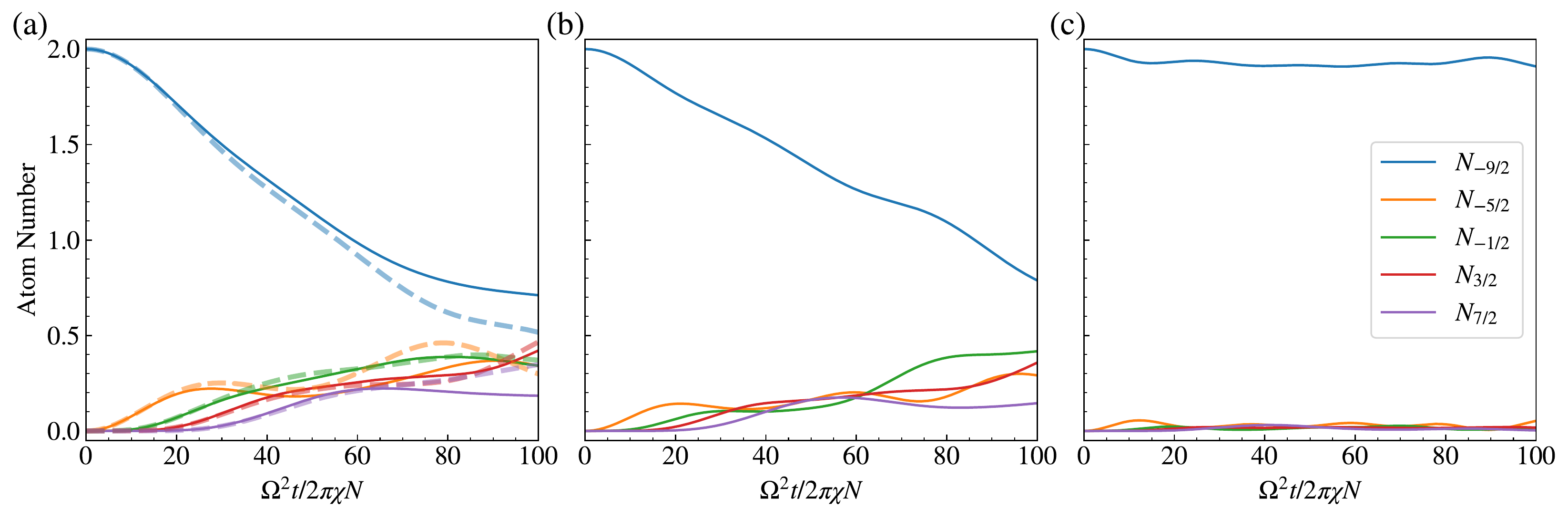}
    \caption{Effect of a finite magnetic field in the short-time dynamics for the case of ${}^{87}$Sr atoms. The initial state is $|g_{-9/2}\rangle^{\otimes(N/2)}|g_{9/2}\rangle^{\otimes(N/2)}$, and  $N=4$. We set $\chi N=2\pi\times 1$MHz, $\Delta_A=-3\chi N$, $\Delta_B=4.1\chi N$. The solid lines are calculated based on Eq.~(\ref{eq:zeeman}), with magnetic field (a) $B=0.12$G ($\delta_e=0.01\chi N$), (b) $B=0.59$G ($\delta_e=0.05\chi N$), and (c) $B=1.77$G ($\delta_e=0.15\chi N$).
    The dashed lines in (a) are the ideal case with zero magnetic field.}
    \label{fig:mag}
\end{figure*}

Although we have assumed a vanishing magnetic field in our discussions, our protocol is relatively robust against the magnetic field along the quantization axis ($\hat{z}$), because the effective ground state Hamiltonian [Eq.~(\ref{eq:heff})] commutes with linear Zeeman shifts on the ground manifold. 
The only constraint for the maximum tolerable magnetic field is the Zeeman shifts of the excited manifold. They should be small compared to $\chi N$, otherwise the many-body structure of the atomic excited states will change and the Zeeman splitting  must be included in the derivation of the effective Hamiltonian. 
We benchmark the effect of magnetic field using the following Hamiltonian,
\begin{equation}
    \hat{H}=\hat{H}_{\mathrm{aa+d}}+\hat{H}_Z, \quad \hat{H}_Z/\hbar=\delta_e\hat{F}_e^z+\delta_g\hat{F}_g^z,\\
    \label{eq:zeeman}
\end{equation}
where $\hat{H}_{\mathrm{aa+d}}$ is defined in Eq.~(\ref{eq:aad}), $\delta_e=\mathcal{G}_{F=9/2,{}^3P_1}\mu_BB$, and $\delta_g=\mathcal{G}_{F=9/2,{}^1S_0}\mu_BB$ with $\mu_B$ the Bohr magneton, and $B$ the magnetic field. Here, $\mathcal{G}_{F=9/2,{}^3P_1}=2/33$, $\mathcal{G}_{F=9/2,{}^1S_0}=-1.3\times 10^{-4}$ are the Land\'{e} g-factors for ${}^{87}$Sr atoms \cite{Boyd2007}.  
We have calculated  the exact short-time dynamics for $4$ atoms using Trotterization, which are shown as solid lines in Fig.~\ref{fig:mag}.
For the case of $0.1$G magnetic field [see Fig.~\ref{fig:mag}(a)], the Hamiltonian dynamics agrees with the case of zero magnetic field at short time.  
Even with a $0.6$G magnetic field [see Fig.~\ref{fig:mag}(b)], it is still possible to observe similar dynamics generated by correlated hopping processes.
A larger  magnetic field can  be problematic since it leads to a significant suppression of the correlated hopping processes [see Fig.~\ref{fig:mag}(c)].

\bibliography{reference}